\def\Journal#1#2#3#4{{#1} {#2} (#4) #3 }

\def\PLA{{\em Phys. Lett.} A}

\def\PREP{\em Phys. Rep.}
\def\PRA{{\em Phys. Rev.} A}
\def\PRD{{\em Phys. Rev.} D}

\def\P{\vec P}

\newcommand{\be}{\begin{equation}}
\newcommand{\ee}{\end{equation}}
\newcommand{\beq}{\begin{equation}}
\newcommand{\eeq}{\end{equation}}
\newcommand{\bea}{\begin{eqnarray}}
\newcommand{\eea}{\end{eqnarray}}

\newcommand{\J} {\Journal}
\newcommand{\qb}{\emph{q}\textsc{Bounce}}
\documentclass[12pt]{iopart}
\usepackage[english]{babel}
\usepackage{epsfig}
\usepackage{subfigure}
\usepackage{graphicx}
\usepackage{iopams}
\usepackage{ifpdf}

\begin{document}

\title{Gravitation and \\quantum interference experiments with neutrons}
\author{Hartmut Abele and Helmut Leeb}

\address{Atominstitut - TU Wien, Stadionallee 2, 1020 WIEN, Austria}
\ead{abele@ati.ac.at}
\begin{abstract}
This article describes gravity experiments, where the outcome depends upon
both the gravitational acceleration $g$ and the Planck constant $\hbar$. We focus on the work done with an elementary particle, the neutron.

\end{abstract}

\tableofcontents
\title[Gravitation and quantum interference experiments with neutrons]{}
\vspace{2pc}

\section{Introduction}

This article describes gravity tests with neutrons. We start with a short note on the first free-fall experiment with neutrons in the classical regime.  Next we consider experiments, where the outcome depends upon
both the gravitational acceleration $g$ and the Planck constant $h$.

In 1975 Colella, Overhauser and Werner demonstrated in their pioneering work~\cite{Colella75} a gravitationally induced phase shift. This
neutron interferometer experiment was carried out at the 2MW
University of Michigan Reactor and the signal is based on the
interference between coherently split and separated neutron de
Broglie waves in the gravity potential.

An other idea is to explore a unique system consisting of a single particle, a neutron falling in the gravity potential of the Earth, and a massive object, a mirror, where the neutron bounces off. The task is to study the dynamics of such a quantum bouncing ball, i.e. a measurement of the time evolution of a coherent superposition of quantum states performing
quantum reflections~\cite{Jenke09,Abele09}. In 2002, an experiment showed that it is possible to populate discrete energy levels in the gravitational field of the earth with
neutrons~\cite{Nesvizhevsky02}. Successor experiments are performed by the \qb-collaboration and the GRANIT-collaboration.

The last point is a measurement of the energy splitting between energy eigenstates with a resonance technique~\cite{Jenke2011}. A novelty of this work is the fact that - in contrast to previous resonance methods -
the quantum mechanical transition is mechanically driven without any direct coupling
of an electromagnetic charge or moment to an electromagnetic potential. Transitions between gravitational quantum states are observed, when a Schr\"odinger-wave packet of an ultra-cold neutron couples to the modulation of a hard surface as the driving force. Such experiments operate on an energy scale of
pico-eVs and can usefully be employed in measurements of fundamental
constants~\cite{Durstberger2011} and in a search for non-Newtonian gravity~\cite{Abele2010}.

Gravity tests with neutrons as quantum objects or within the classical limit are reviewed in~\cite{AbelePPNP} and this work is an update. The large field of neutron optics and neutron interferometry has been omitted, because it can be found in ~\cite{Rauch2011}. Neutron
interferometry experiments are also discussed in a review on ``Neutron
Interferometry"~\cite{Rauch00}. Much of the material on quantum
interference phenomena has been covered there.

\subsection{\it Neutron sources, neutron interferometers, and neutron mirrors\label{sec:nproduction}}
Neutrons are typically produced in a research reactor or a
spallation source. Free neutrons cover kinetic energies across the whole energy range from several 100 MeV
to $10^{20}$ times less at the pico-eV energy scale. The various fluxes are conveniently
defined~\cite{Byrne93} as fast ($E_n$ $>$ 1 MeV), of intermediate
energy (1 MeV $\geq$ $E_n$ $>$ 1 eV), or slow ($E_n$ $<$ 1 eV). Here
several subgroups are classified as epithermal (1eV $\geq$ $E_n$ $>$
0.025 eV), thermal ($\simeq\,$0.025 eV), cold (0.025 eV $\geq$ $E_n$
$\geq\, 0.5\, \mu$eV), very cold (0.5 $\mu$eV $>$ $E_n$ $\geq$ 100
neV) and ultracold (UCN) ($<$ 100 neV)~\cite{Byrne93}. From the experimental point of view, ideas have been developed for techniques that
would produce a 1000-fold increase in the UCN counting rate due to a
higher UCN phase space density. New UCN sources are therefore planned or under construction at many
cites, including facilities at FRM-II, ILL, KEK, LANL, Mainz, North
Carolina State, PNPI, PSI and Vienna.

\begin{table}[t]
\caption{from hot to ultracold: neutron energy, temperature,
velocity and wavelength distributions}
\begin{center}
\renewcommand{\arraystretch}{1.4}
\setlength\tabcolsep{5pt}
\begin{tabular}{lllllll}
\hline\noalign{\smallskip} & ${\mathrm {fission}}$ & ${\mathrm
{thermal}}$ & ${\mathrm {cold}}$ & ${\mathrm
{ultracold}}$&${\mathrm {gravity}}$\\
& ${\mathrm {neutrons}}$ &
${\mathrm {neutrons}}$ & ${\mathrm {neutrons}}$ & ${\mathrm {neutrons}}$&${\mathrm {experiment}}$\\
\hline\noalign{\smallskip}
Energy & 2 MeV & 25 meV & 3 meV & $<$ 100 neV & 1.4 peV (E$_{\perp}$)  \\
Temperature & 10$^{10}$ K & 300 K & 40 K & $\sim$ 1 mK & - \\
Velocity & 10$^7$ m/s & 2200 m/s & 800 m/s & $\sim$ 5 m/s & $v_{\perp}\sim$ 2 cm/s\\
Wavelength & & 0.18 nm & 0.5 nm & $\sim$ 80 nm& &\\ \hline
\end{tabular}
\end{center}
\label{apptab1b}
\end{table}

\begin{figure}[tb]
\subfigure{\hspace{2.5cm}\epsfig{file=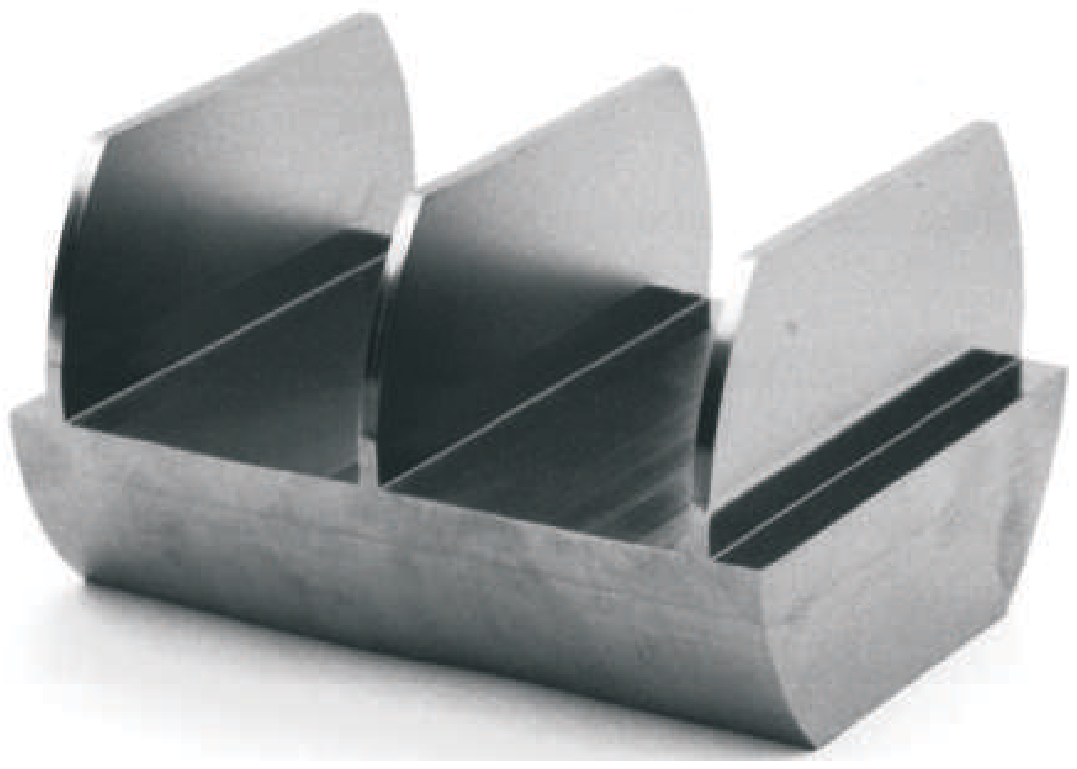,scale=.36}}\hspace{3.0cm}\epsfig{file=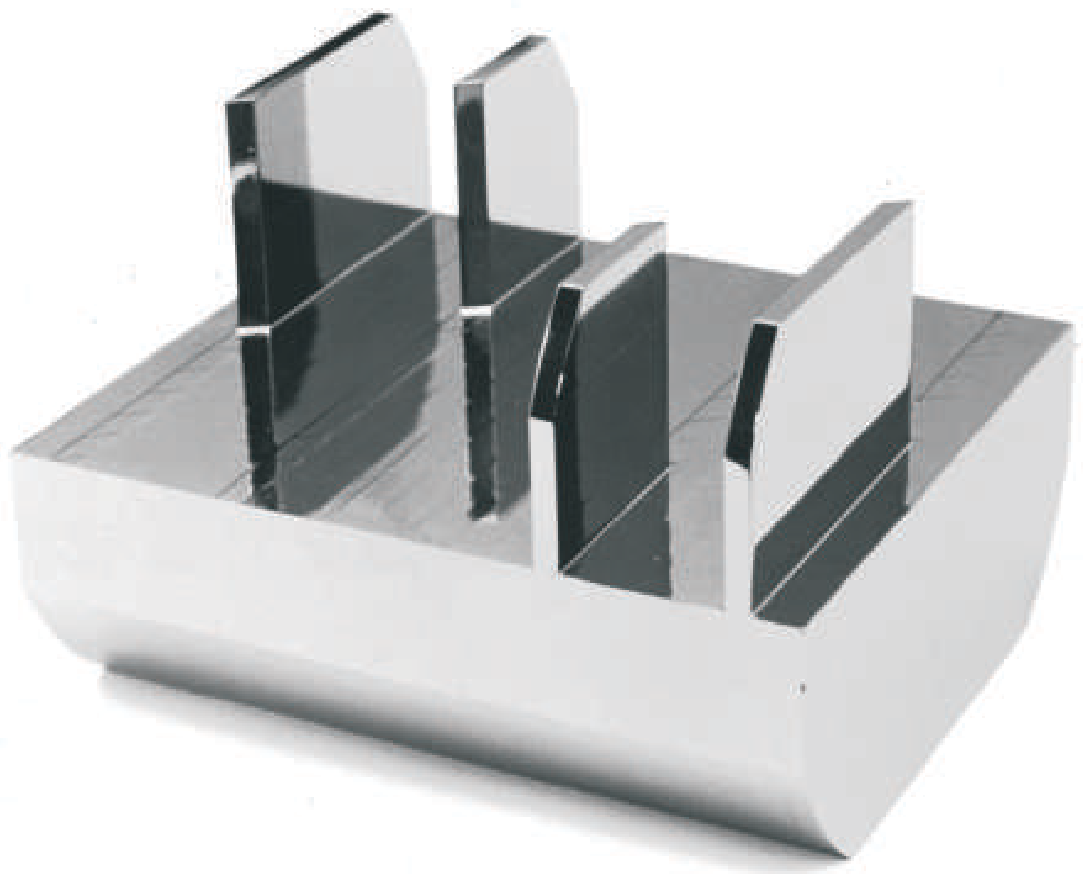,scale=.36}
\caption{A Photograph of the symmetric (left) and skew-symmetric
(right) interferometer used in~\cite{Littrell97}. The blades of the
symmetric interferometer are 3.077 mm thick and 50.404 mm apart. The
dimensions of the skew-symmetric interferometer are 16.172 mm and
49.449 mm. The blade thickness is 2.621 mm. \label{fig:photo}}
\end{figure}

Neutrons propagate in condensed matter in a manner similar to the
propagation of light, but with a neutron refractive index of less
than unity. Interferometers are based on the atom arrangement in silicon crystals and beams are split coherently in ordinary or momentum space for interferometry experiments, where the relative phase becomes a measurable quantity. In this case nuclear, magnetic and gravitational phase shifts can be applied and measured precisely. A monolithic design and a stable environment provide the parallelity of the lattice planes throughout the interferometer. Fig.~\ref{fig:photo} shows interferometers used in gravity
experiments, a symmetric and a skew-symmetric one in order to
separate and identify systematic discrepancies that depend on the
interferometer or the mounting. For each wavelength, the phase
difference $\Delta(\lambda)$ was obtained using a technique
called the phase rotator interferogram technique. A phase rotator is
a 2-mm thick aluminium phase flag, which is placed across both beams
and rotated.
\pdffalse

The surface of matter
constitutes a potential step of height $V$. Neutrons with
transversal energy $E_{\perp}$ $<$ $V$ will be totally reflected.
Following a pragmatic definition, ultracold neutrons (UCN) are
neutrons that, in contrast to faster neutrons, are retro-reflected
from surfaces at all angles of incidence. When the surface roughness
of the mirror is small enough, the UCN reflection is specular.

This
feature makes it possible to build simple and efficient
retroreflectors. Such a neutron mirror makes use of the strong
interaction between nuclei and UCN, resulting in an
effective repulsive force. The potential may also be based on the
gradient of the magnetic dipole interaction. Retroreflectors for atoms have used
the electric dipole force in an evanescent light
wave~\cite{Aminoff93,Kasevich90} or are based on the gradient of the
magnetic dipole interaction, which has the advantage of not
requiring a laser~\cite{Roach95}.

\section{Gravity experiments with neutrons within the classical limit}
In the earth's gravitational field,
neutrons fall with an acceleration equal to the local value
$g$~\cite{McReynolds51}.  The free fall does not depend on the sign of neutron's vertical spin
component~\cite{Dabbs65}. The studies provide
evidence in support of the weak equivalence principle of equality of
inertial mass $m_i$ and gravitational mass $m_g$. The results
obtained by Koester et al. confirm that $m_i/m_g$ is equal to
unity to an accuracy of 3 $\times$
10$^{-4}$~\cite{Koester76,Sears82} and is a consequence of
classical mechanics and has been demonstrated by verifying that
neutrons fall parabolically on trajectories in the earth's
gravitational field.

\section{Gravitation and neutron-interferometry\label{sec:GravInd}}
R. Colella, A. W. Overhauser and S. A. Werner (COW) observed a
``quantum-mechanical phase shift of neutrons caused by their
interaction with Earth's gravitational field"~\cite{Colella75}. The signal is based on the
interference between coherently split and separated neutron de
Broglie waves in the gravity potential. The amplitudes are divided
by dynamical Bragg diffraction from perfect silicon crystals. Such
interferometers were originally developed for X-rays and then adapted for
thermal neutrons by Rauch et al. in 1974~\cite{Rauch74}. This
neutron interferometer experiment was carried out at the 2MW
University of Michigan Reactor.

\begin{figure}
\hspace{3cm}\epsfig{file=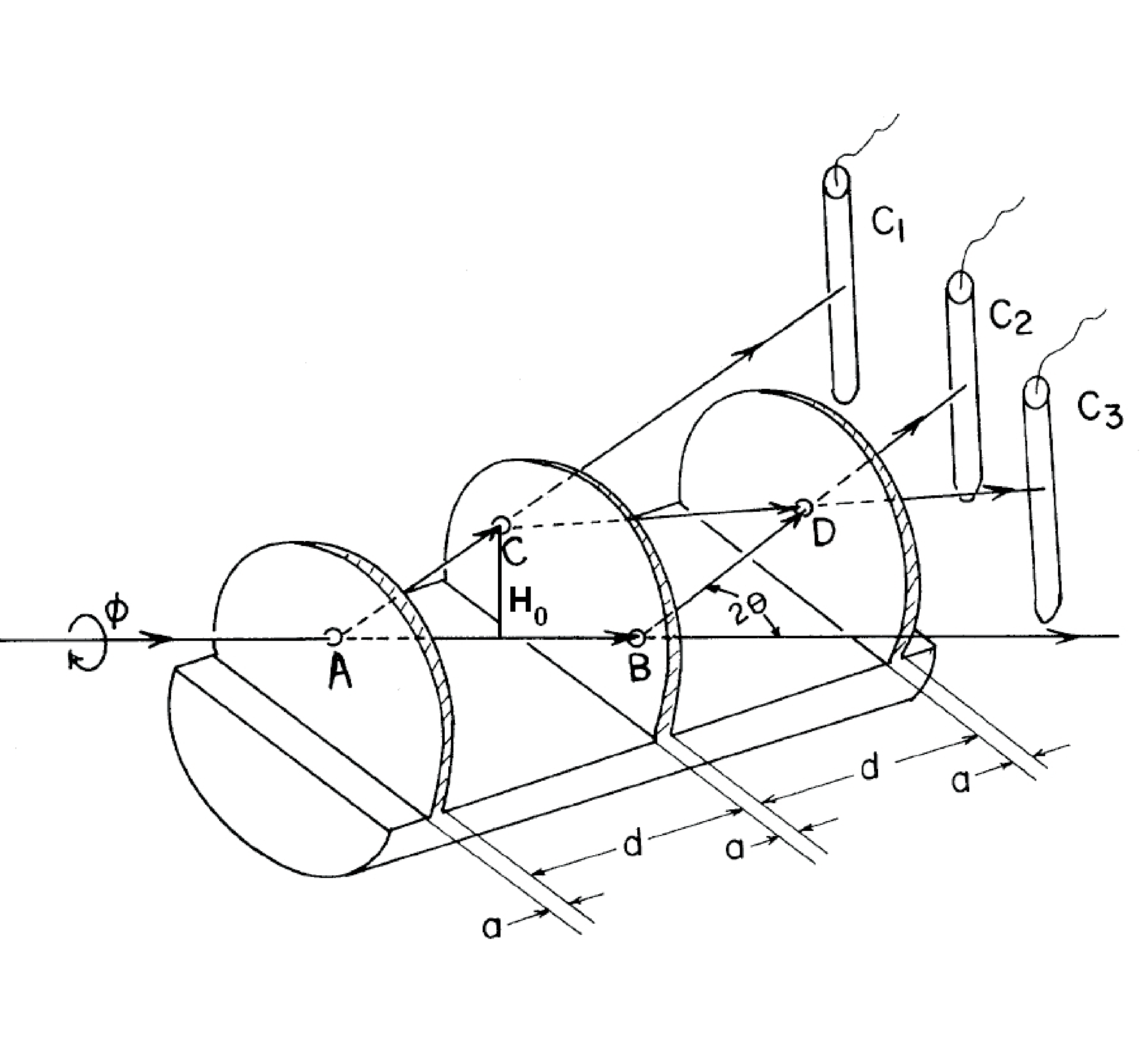,scale=.3}
\caption{Sketch of
the interferometer used in the COW experiment~\cite{Colella75}.
\label{cow_setup}}
\end{figure}

\begin{figure}[tb]
\subfigure{\hspace{2cm}\epsfig{file=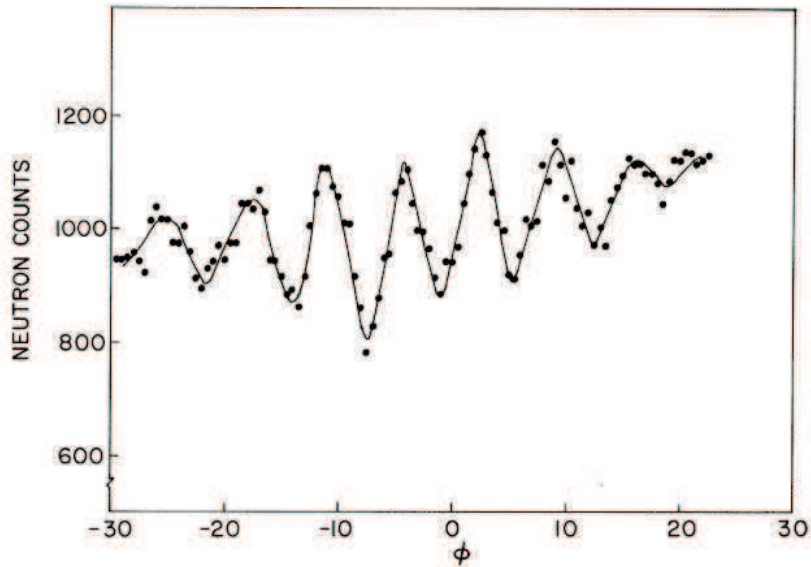,scale=.69}}
\caption{Gravitationally induced quantum interferogram~\cite{Colella75}.}
\label{cow_result}
\end{figure}

The following description of the COW experiment, shown in Fig. \ref{cow_setup}, follows in part~\cite{Rauch00,AbelePPNP}. A
monochromatic neutron beam with wavelength $\lambda$ enters a
standard triple-plate interferometer along a horizontal line
with momentum $p_0$ = $\hbar k_0$. The neutron is divided
into two wavepackets at point $A$ following sub-beam paths $ABD$ and $ACD$. The
interferometer is turned around the incident beam direction by an
angle $\phi$ maintaining the Bragg condition. The neutron
wavepackets mix and recombine in the third crystal plate at point
$D$, which has a higher gravitational potential above the Earth than
the entry point $A$ by a size $m_n g H(\phi) = m_n g
H_0\mathrm{sin}(\phi)$, where $m_n$ is the neutron mass and
$g$ is the acceleration of the earth. The sum of the kinetic energy
and the potential energy is constant, \be E_0 =
\frac{\hbar^2k_0^2}{2m_n}=\frac{\hbar^2k^2}{2m_n} + m_ngH(\phi), \ee
but the difference in height implies that the momentum $p$ = $\hbar
k$ on path $CD$ is less than the momentum $p_0$ = $\hbar k_0$ on
path $AB$. Fig. \ref{cow_result} shows the difference in the count
rate in counter C2 and C3 as a function of angle $\phi$.

\pdffalse
The gravitationally induced quantum interferogram is due to the
phase difference of the neutron traversing path $ACD$ relative to
path $ABD$
\begin{equation}
\begin{array}{rcl}
\Delta \Phi _{COW}& =& \Phi _{ACD} - \Phi _{ABD} \\
&=&\Delta kS \\
&\simeq&-q_{COW}\sin \phi,
\label{eq:cowfreq}\end{array}
\end{equation}
where $\Delta k$ =$k - k_0$ is the difference in the wave numbers, $S$ the path length of the
segments AB and CD, and $q_{COW}= 2\pi \lambda \frac{m_n^2}{h^2}gA_0$.  Here, $A_0$ = $H_0 S$ is the area of the parallelogram
enclosed by the beam path.

The COW measurement differs slightly from the expected value of
Eq.~\ref{eq:cowfreq} due to additional phaseshifts caused by the
bending of the interferometer under its weight and by the earth's
rotation, the Sagnac effect. There are small dynamical diffraction effects,
too, shifting the central frequency of the oscillations from
$q_{COW}$ to \be q_{grav} = q_{COW}(1+\epsilon), \ee where the
correction factor $\epsilon$ of several percent depends on the thickness-to-distance ratio of the interferometer plates. The total phase shift
of tilt angle $\phi$ practically originates from three
terms~\cite{Rauch00}, \be
\begin{array}{rcl}
 \Delta \Phi(\phi) &=&
\Delta \Phi_{grav}(\phi) + \Delta \Phi_{bend}(\phi) + \Delta \Phi_{Sagnac}(\phi) \\
&=&-q_{grav}\sin\phi -q_{bend}\sin\phi + q_{Sagnac}\cos\phi \\
&=&q \sin(\phi - \phi_0), \\
\end{array}
\ee where the frequency of the oscillation with regard to the tilt angle is given by \be q =
\left[(q_{grav} + q_{bend})^2 + (q_{Sagnac})^2 \right]^{1/2} \ee and
\be \phi_0 = \tan^{-1} \left[\frac{q_{Sagnac}}{q_{grav} +
q_{bend}} \right] . \ee A comparison of gravitationally induced
quantum interference experiments is given in Table~\ref{tab:cowexp},
providing information on the interferometer parameters, the neutron
wavelength $\lambda$, and the interferogram oscillations $q_{COW}$.
In the following years, these systematic effects have been studied
carefully.  In the experiment of Ref.~\cite{Werner88}, the bending
was separately measured with X-rays and Eq.~\ref{eq:cowfreq} reads
numerically \be
\begin{array}{rcl}
q_{grav} &=& \left(q_{exp}^2 - q_{Sagnac}^2\right)^{1/2} - q_{bend}\\
&=&
(60.12^2 - 1.45^2)^{1/2} - 1.42\, \mathrm{rad} \\
&=& 58.72 \pm 0.03 \,\mathrm{rad}.
\end{array} \ee The theoretical prediction $q_{grav} = 59.2 \pm 0.1 $rad is 0.8 \% higher.

The gravitationally induced phase shift is proportional to
$\lambda$, and the phase shift due to bending is proportional to
$\lambda^{-1}$. Therefore, a simultaneous measurement with two
neutron wavelengths can be used to determine both contributions.

In an experiment by Littrell et al.~\cite{Littrell97}, nearly
harmonic pairs of neutron wavelengths are used to measure and
compensate for effects due to the distortion of the interferometer
as it is tilted about the incident beam direction.
Fig.~\ref{fig:photo} shows the interferometers used in this
experiment, a symmetric and a skew-symmetric one in order to
separate and identify systematic discrepancies that depend on the
interferometer or the mounting. For each wavelength, the phase
difference $\Delta(\lambda,\phi)$ was obtained using a phase rotator, which is placed across both beams
and rotated. A series of phase rotator
scans were taken for various values of $\phi$ using the wavelengths
0.21440 nm and 0.10780 nm. The corresponding Bragg angles $\Theta_B$
are 34.15$^\circ$ and 33.94$^\circ$. The Si(220) or Si(440) Bragg
reflection is used. The phase advances by almost the same amount
with each step and nearly twice as much for the long wavelength as
for the shorter wavelength. Previously, the neutron Sagnac phase
shift due to the Earth's rotation has been
measured~\cite{Werner86,Atwood84,Staudenmann80} and has shown to be
in agreement with the theory of the order of a few percent. This
phase shift is small for the tilt angles spanned in this experiment
and has been subtracted from the data. A historical summary of
gravitationally-induced quantum interference
experiments~\cite{Werner88,Littrell97,Staudenmann80,Zouw00} can be
found in~\cite{Rauch00}. The most precise measurements by Littrell et
al.~\cite{Littrell97} with perfect silicon crystal interferometers
show a discrepancy with the theoretical value of the order of 1\%.

The experiment by van der Zouw et al.~\cite{Zouw00} uses very cold
neutrons with a mean wavelength of $\lambda$ = 10 nm (a velocity of
40 m/s) and phase gratings as the beam splitting mechanism. The
interferometer has a length of up to several meters. The results are
consistent with the theory within the measurement accuracy of
1\%~\cite{Zouw00}. The experiments of Refs.~\cite{Staudenmann80,Littrell97} were done at the 10 MW  University of Missouri Research Reactor.

\begin{table}[t]
\begin{center}
\begin{minipage}[t]{14.5 cm}
\hspace{-2cm}\caption{History of gravity-induced interference experiments with
symmetric (sym.) and skew-symmetric silicon interferometers. The
restricted (rest.) range data means that the tilt angle $|\phi|=
11^\circ$. The two wavelengths of~\cite{Littrell97} are diffracted
by the (220) or (440) lattice planes. The table is based
on~\cite{Rauch00}.} \label{tab:cowexp}
\end{minipage}
{\small
\item[]
\begin{tabular}{c|lccccc} \hline

 Ref. & Inter- & $\lambda$[nm]& A$_0$ [cm$^2]$&$q_{COW}$ (theory)&$q_{COW}$
 (exp)&Agreement
                 \\
   & ferometer& & &        [rad] &[rad]& with theory          \\ \hline
 \cite{Colella75}               &Sym.\#1  &1.445(2)  &10.52(2)&59.8(1)&54.3(2.0)
 &12\%\\
 \cite{Staudenmann80}&Sym.\#2&1.419(2)&10152(4)&56.7(1)&54.2(1)&4.4\%   \\
  & &1.060(2)&7,332(4)&30.6(1)&28.4(1)&7.3\% \\
 \cite{Werner88}&Sym.
 \#2&1.417(1)&10.132(4)&56.50(5)&56.03(3)&0.8\% \\
\cite{Littrell97}&Skew-sym.&&&&&\\
 (440)&full range&1.078(6)&12.016(3)&50.97(5)&49.45(5)&3.0\%\\
 &rest. range&1.078(6)&12.016(3)&50.97(5)&50.18(5)&1.5\% \\
(220)&full range
&2.1440(4)&11.921(3)&100.57(10)&97.58(10)&3.0\%\\
&rest. range&2.1440(4)&11.921(3)&100.57(10)&99.02(10)&1.5\%\\
 \cite{Littrell97}&Large Sym. &&&&&\\
(440) &full range&1.8796(10)&30.26(1)&223.80(10)&223.38(30)&0.6\%\\
(220)&rest. range&1.8796(10)&30.26(1)&223.80(10)&221.85(30)&0.9\%\\
\hline

\end{tabular}
}
\end{center}

\end{table}

The results are remarkable in some respects. First of all, they demonstrate the
validity of quantum theory in a gravitational potential on the 1\% level. The
gravity-induced interference is purely quantum mechanical, because
the phase shift is proportional to the wavelength $\lambda$ and
depends explicitly on Planck's constant $\hbar$ but the interference
pattern disappears as $\hbar \rightarrow$ 0. The
effect depends on $(m_n/\hbar)^2$ and the
experiments test the equivalence principle.

It has been noted, that the gravity-induced quantum interference has a
classical origin, due to the time delay of a classical particle
experienced in a gravitational background field, and classical
light waves also undergo a phase shift in traversing a gravitational
field~\cite{Mannheim98}. Cohen and Mashhoon~\cite{Cohen93} already
derived this result with the argument that the index of refraction
$n$ in the exterior field of a spherically symmetric distribution of
matter is only a function of the isotropic radial coordinate of the
exterior Schwarzschild geometry, which is asymptotically flat.

In recent years, ultracold atoms
have been used to perform high-precision measurements with
atom-interferometers. This technique has been reviewed in~\cite{Pritchard2009}. The Stanford
group~\cite{Peters99} measured $g$ with a resolution of
$\Delta g/g$ = 1 $\times$ 10$^{-10}$ after two days of integration time. The result has been reinterpreted as a precise measurement of the gravitational redshift by the interference of matter waves ~\cite{Mueller2010}. The interpretation has been questioned~\cite{Wolf2010,Wolf2011}, see also ~\cite{Mueller2010a}.

\section{The quantum bouncing ball\label{sec:boundqs}}
Above a mirror, the gravity potential leads to discrete energy levels
of a bouncing massive particle. The corresponding quantum mechanical motion of
a massive particle in the gravitational field has been named the quantum
bouncer~\cite{Abele09,Gibbs,Rosu,qqb}. The discrete energy levels occur due to
the combined confinement of the matter waves by the mirror and the
gavitational field. For neutrons the lowest discrete states are in the range
of several peV, opening the way to a new technique for gravity experiments and
measurements of fundamental properties.

The quantum mechanical description of ultracold neutrons of mass $m_n$ moving
in the gravitational field above a mirror is essentially an one-dimensional
problem. The corresponding gravitational potential is usually given in linear
form by $m_ngz$, where $g$ is the gravitational accelaration and $z$ is the
distance above the mirror, respectively. The mirror, frequently made of glass,
with its surface at $z=0$ is represented by a constant potential $V_{mirror}$
for $z\leq.0$. The potential $V_{mirror}$ is essentially real because of the
small absorption cross section of glass and with about 100 neV large compared
to the neutron energy $E_{\perp}$ perpendicular to the surface of the
mirror. Therefore it is well justified to assume the mirror as a hard boundary for
neutrons at $z=0$.

The time-dependent Schr\"odinger equation for the neutron quantum bouncer is
given by
\be
-\frac{\hbar^2}{2m_n}\frac{\partial^2\Psi}{\partial z^2}+m_ngz\Psi =
\imath\hbar\frac{\partial \Psi}{\partial t}\quad \mbox{for}\quad z > 0\, ,
\label{qb-tdeq}
\ee
where the mirror surface is at $z=0$. Assuming an infinite mirror potential
$V_{mirror}$ implies that $\Psi (z,t)$ vanishes at all times at the mirror
surface. The solution of Eq. (\ref{qb-tdeq}) is given by a superposition
energy eigenstates
\be
\Psi(z,t) = \sum C_n e^{-\imath E_n t /\hbar}\psi_n(z)\, .
\label{superposition}
\ee
The coefficients $C_n$ are determined from the initial condition $\Psi(z,0)$,
and the eigenfunctions $\psi_n$ are the solutions of the time-independent
Schr\"odinger equation
\be
\left\{ -\frac{\hbar^2}{2m_n}\frac{\partial^2}{\partial z^2}
+ V(z)\right\}\psi_n = E_n\psi_n
\label{qb-steq}
\ee
with
\be
V(z)=\left\{\begin{array}{r@{\quad:\quad}l} m_ngz & z\ge 0\\
\infty & z<0 \end{array}\right. \, .
\label{eq:sgl}
\ee
This equation describes the neutron in the gravitational field above a mirror
at rest. It is convenient to scale (\ref{qb-steq}) with the characteristic
gravitational length scale~\cite{Wallis} of the bouncing neutron
\be
z_0 =\left(\frac{\hbar^2}{2m_n^2g}\right)^{1/3} = 5.87\,\mathrm{\mu m}\, .
\ee
and the corresponding characteristic gravitational energy scale is
\be E_0 = \left(\frac{\hbar^2m_ng^2}{2}\right)^{1/3} = 0.602\, \mathrm{peV}\,.
\label{eq:energy}
\ee
Thus Eq. (\ref{qb-steq}) takes the concise form
\begin{equation}
\psi^{\prime\prime}(\zeta)-(\zeta-\zeta_E)\psi(\zeta)=0, \label{eq:zeta}
\end{equation}
where $\zeta=z/z_0$, $\zeta_E=z_E/z_0$, and $z_E = E/mg$.

\begin{figure}[tb]
\hspace{2cm}\epsfig{file=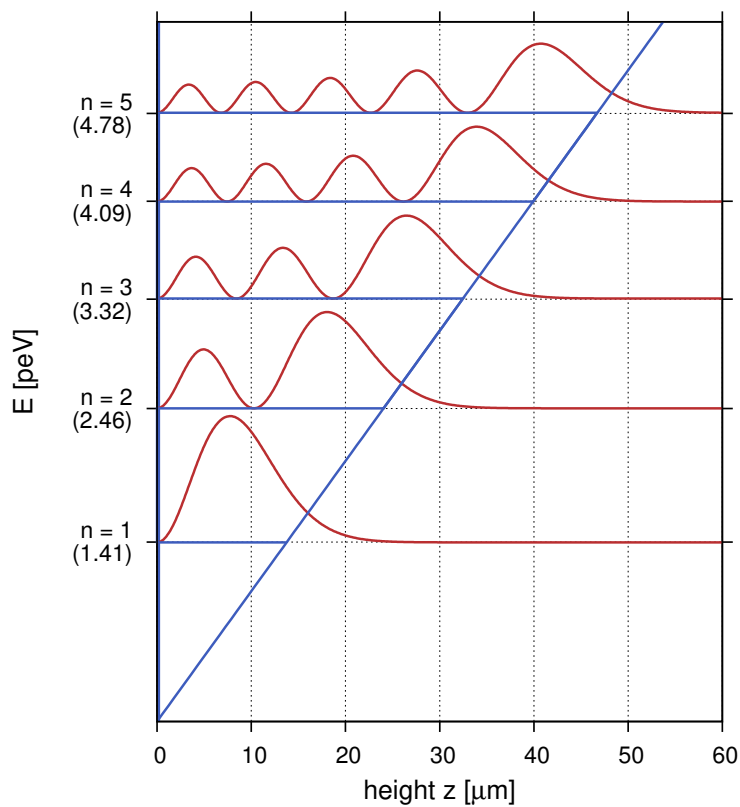,scale=1.25}
\caption{Energy eigenvalues (blue) and neutron density distributions (red) for
level one to five.}
\label{fig:states}
\end{figure}

For a neutron above a vertical mirror the linear gravity potential leads to
the following discrete energy eigenstates: The lowest energy eigenvalues
$E_n$, (n = 1, 2, 3, 4, 5), are 1.41 peV, 2.46 peV, 3.32 peV, 4.09 peV, and
4.78 peV.  The corresponding classical turning points $z_n$ are 13.7 $\mu$m,
24.1 $\mu$m, 32.5 $\mu$m and 40.1 $\mu$m (see table~\ref{picoev}).
The energy levels together with the neutron density distribution are shown in
Fig.~\ref{fig:states}. The idea of observing quantum effects in such a
gravitational cavity was discussed with neutrons~\cite{Lushikov} as well as
with atoms~\cite{Wallis}. The corresponding eigenenergies and associated
classical turning points for states with $n=1$ and $2$ of a neutron, an
electron and several atoms are compared in table~\ref{tab:qstates}.
\begin{table}
\caption{Eigenenergies and classical turning points for neutrons,
atoms and electrons}

\hspace{2.3cm}\begin{tabular}{llllll}
\hline\noalign{\smallskip} & ${\mathrm {Neutron}}$ &
$^{4}{\mathrm {Helium}}$ & $^{85}{\mathrm {Rubidium}}$ & $^{133}{\mathrm {Cesium}}$&${\mathrm {Electron}}$\\
\noalign{\smallskip} \hline \noalign{\smallskip}
$E_1\; [peV]$ & 1.4 & 2.3 & 6.2 & 7.2 & 0.12 \\
$E_2\; [peV]$ & 2.5 & 3.9 & 11.0 & 12.7 & 0.20 \\
$z_1\; [\mu\mathrm{m}]$ & 13.7 & 5.5 & 0.7 & 0.5& 2061\\
$z_2\; [\mu\mathrm{m}]$ & 24.0& 9.5 & 1.2 & 0.9 & 3604\\
\hline\label{picoev}\label{tab:qstates}
\end{tabular}
\end{table}

As Gea-Banacloche~\cite{qqb} pointed out, the eigenfunctions for this problem
are {\em pieces} of the same Airy function in the sense that they are shifted in
each case in order to be zero at $z$ = 0 and cut for $z<0$. Secondly, the
wavelength of the oscillations decreases towards the bottom at small $z$. This is in
accordance with the de Broglie relation $\lambda = h/mv$, since the velocity
of a classical particle is greater there.

The \qb-experiment has been focussed on the Quantum Bouncing Ball i.e. a
measurement of the time evolution of the Schr\"odinger wave function of a
neutron bouncing above a mirror~\cite{Abele09}.
The experiment at the Institut Laue-Langevin, Grenoble, has been performed in
the following way~\cite{Jenke09}: Neutrons are taken from the ultra-cold neutron installation
PF2. A narrow beam of ultra-cold neutrons is prepared with an adjustable
horizontal velocity range between 5 m/s and 12 m/s. Fig.~\ref{fig:setup},
left, shows a sketch of the experiment. At the entrance of the  experiment, a
collimator absorber system limits the transversal velocity to an energy in the
pico-eV range, see table~\ref{picoev}. The experiment itself is mounted on a
polished plane granite stone with an active and passive antivibration table
underneath. This stone is leveled using piezo translators. Inclinometers
together with the piezo translators in a closed loop circuit guarantee
leveling with a precision better than 1 $\mu$rad. A solid block with
dimensions 10 cm $\times$ 10 cm $\times$ 3 cm composed of optical glass serves
as a mirror for neutron reflection. The neutrons see a surface that is
essentially flat. An absorber/scatterer, a rough mirror coated with a neutron
absorbing alloy, is placed above the first mirror at a certain height in order
to select different quantum states. The neutrons are guided through this
mirror-absorber-system in such a way that they are in first few quantum
states. Neutrons being in higher, unwanted states are scattered out of this
system. The neutron loss mechanism itself is described at length
in~\cite{Westphal02,Voronin06,Westphal06}. This preparation mechanism can be
simply modeled by a general phenomenological loss rate $\Gamma_n(l)$ of
the n$^{\rm th}$ bound state, which is taken to be proportional to
the probability density of the neutrons at the absorber/scatterer.
\begin{figure}[tb]
\hspace{2cm}\epsfig{file=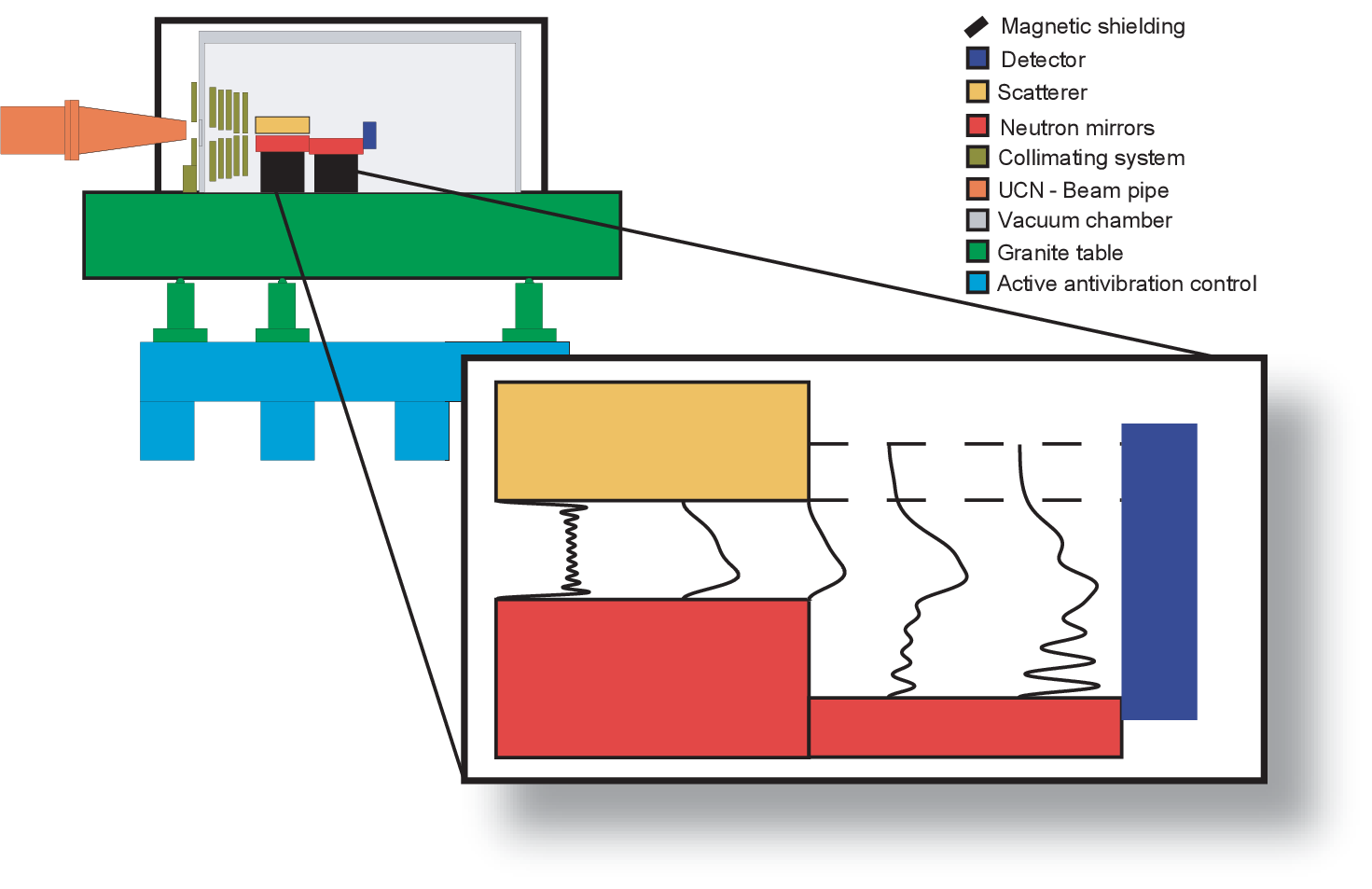,scale=0.7}
\caption{Sketch of the setup. Neutrons are prepared in the lower quantum states and then fall a down a step of 27 $\mu$m.}
\label{fig:setup}
\end{figure}
\begin{figure}[tb]
\hspace{2cm}\epsfig{file=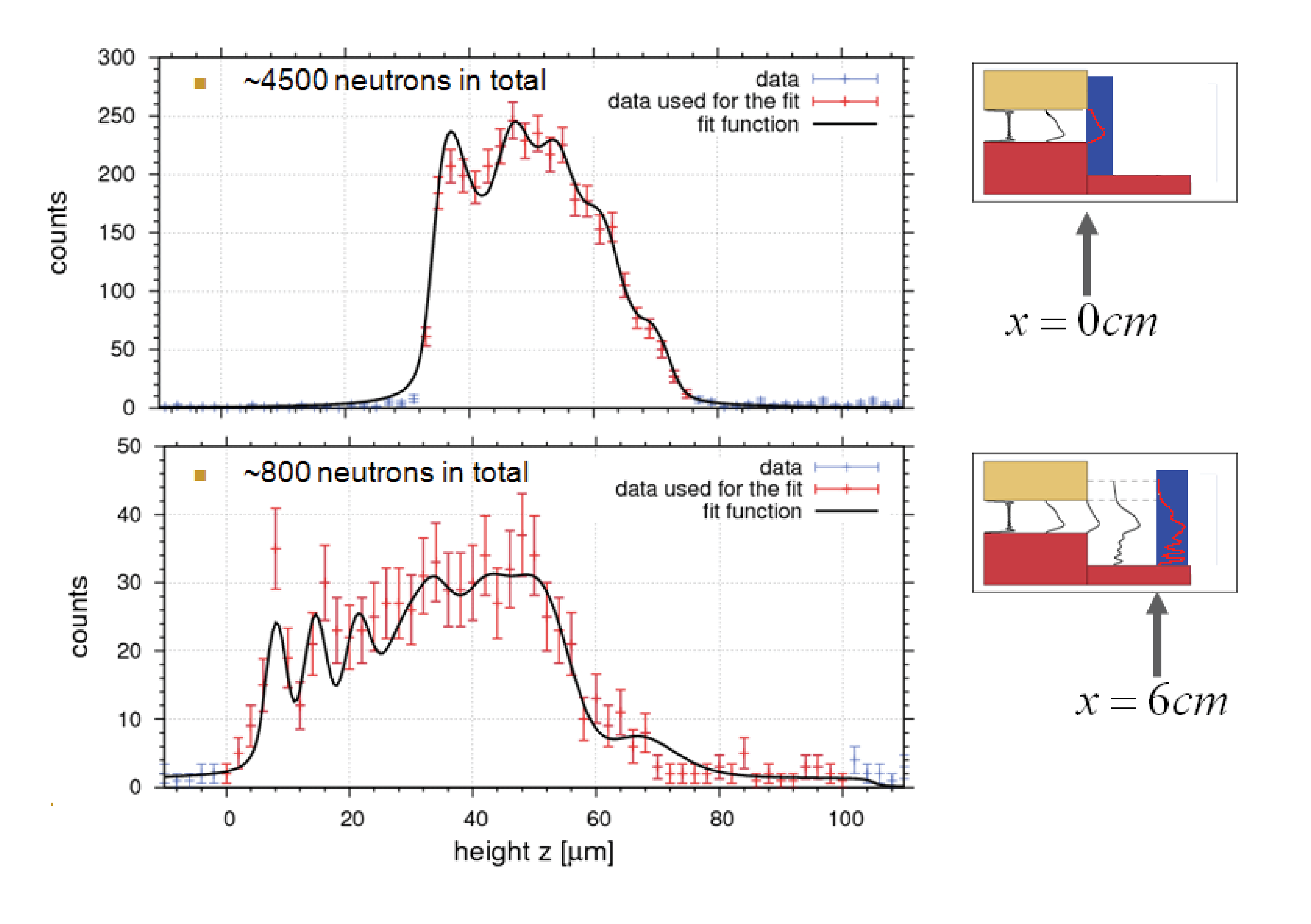,scale=0.4}
\caption{A measurement of the Quantum Bouncer. Upper figure: A fit to the
square of the prepared Schr\"odinger wave functions just at the step ($x$ = 0
cm). Lower figure: Quantum prediction after falling and rebouncing ($x$ = 6
cm). The figure is taken from~\cite{Abele09}.}
\label{fig:setup}
\end{figure}

After this preparation process the quantum state falls down a step of several
$\mu$m. This has been achieved with a second mirror, which is placed after the
first one, shifted down by 27 $\mu$m. The neutron is now in a coherent
superposition of several quantum states. Quantum reflected, the wave-function
shows aspects of quantum interference, as it evolves with time.

The Schr\"odinger-wave packet - the spatial probability distribution - has
been measured at four different positions with a spatial resolution of about
1.5 $\mu$m. Fig.~\ref{fig:setup} shows a first result with CR39 detector
at positions $x$ = 0 cm, $x$ = 6 cm; the quantum fringes are already visible
here. The theory makes use of the independently determined parameters for the
scatterer and has been convoluted with the spatial detector resolution.

The spatial resolution detector has been developed using $^{10}$B-coated
organic substrates (CR39)~\cite{Ruess00}. The detectors operate with an
absorptive layer of $^{10}$Boron on plastic CR39. Neutrons are captured in
this coated Boron layer in a Li-$\alpha$-reaction. Boron has two naturally
occurring and stable isotopes, $^{11}$B (80.1\%) and $^{10}$B (19.9\%). The
$^{10}$B isotope is ideal at capturing thermal neutrons. The
Li-$\alpha$-reaction converts a neutron into a detectable track on CR39. An
etching technique makes the tracks with a length of about 3 $\mu$m
visible. The information is retrieved optically by an automatic scanning
procedure. The microscope is equipped with a micrometer scanning table and a
CCD camera of sufficient quality. A drawback of this procedure is that it is
very time consuming and does not allow for fast feed-back. In addition,
corrections are large and of the order of between 5$\mu$m and 50$\mu$m. An
electronic alternative, which allows an online-access to the data, are modern
CCD or CMOS chips. Again, a thin boron layer will serve as neutron
converter.

A visualization of the neutron motion of a quantum bouncer is very helpful for
the interpretation of the experiment. Therefore a simulation~\cite{RSS2009} of
the quantum bouncer has been set up following the theoretical basis outlined in
Eqs. (\ref{qb-tdeq}) and (\ref{superposition}). The assumed experimental
setup of the animation is equivalent to that of Fig. \ref{fig:setup}, but
with the step height of 40 $\mu$m instead of 27 $\mu$m. At the beginning of
the animation the neutron wave is in the quantum state $n=1$ above a mirror at
height $z=40 \mu$m. Assuming a Gaussian wave packet with a width of 1 $\mu$m
and mean velocity of 6 m/s incident in x-direction passes the edge, after
which it has to be represented as a superposition of many discrete eigenstates
above the mirror at $z=0 \mu$m. In this new represention the largest
contributions are from the states with $n=5,6,7$, but contributions greater
than 1\% can be found up to $n=19$.

\begin{figure}[tb]
\subfigure{\hspace{-1.8cm}\epsfig{file=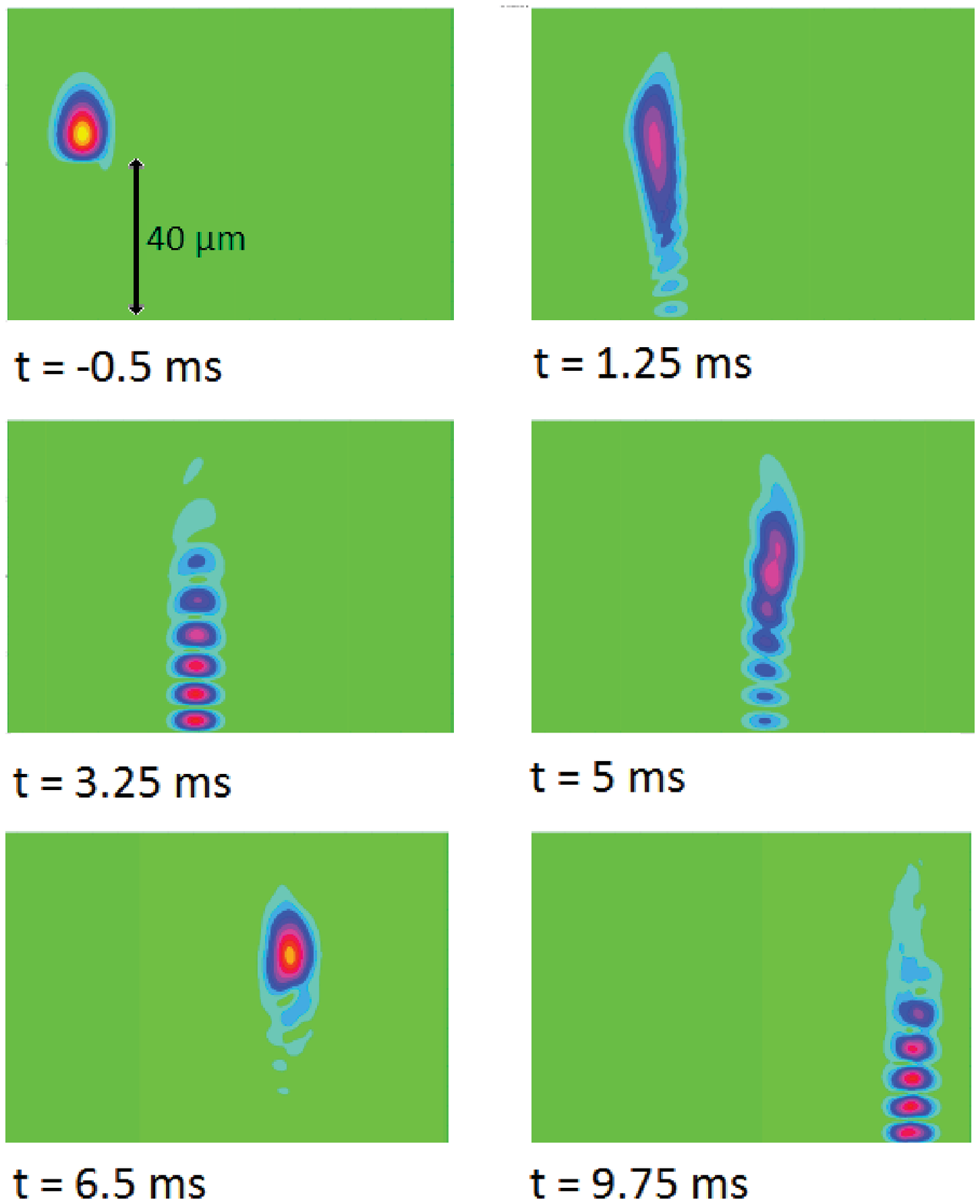,scale=.75}}
\caption{This figure shows a simulation of the quantum bouncing ball at different time steps. A neutron in quantum state $|1>$ is localized at a height of 40 $\mu$m and then falling, reflecting and bouncing back.
\label{fig:qbb3}}
\end{figure}

Fig.~\ref{fig:qbb3} shows the spatial probability distribution with several snapshots of the fall and rise of the wave packet. Just in front of the 40 $\mu$m step at t = -0.5 ms, the Schr\"odinger-wave packet is localized in state $|1>$. At t = 1.25 ms the figure shows an intermediate step during the fall; at t = 3.25 ms the wave packet is performing a quantum reflection; t = 5 ms shows the rise; at t = 6.5 ms the turning point is reached; and at t = 9.75 ms the wave packet is performing a second reflection.

The neutron quantum bouncer exhibits several interesting
quantum phenomena such as collapses and revivals of the wave function. These
are a consequence of the different evolution of the phases of the contributing
quantum states. The expectation value of the time evolution shows a pattern
similar to quantum beats, but is not harmonic due to the $n$-dependence of
energy differences $E_n-E_{n-1}$ in the gravitational potential. Following the
evolution in detail one observes that the first few bounces are clearly
visible and follow the classical motion. As the time goes on, it is not
possible to tell whether the particle is falling down or is going up, and the
expectation value $\langle z \rangle$ of the wave packet remains very close to
the time average of the classical trajectory. Later, the oscillations start
again and the particle bounces again. The revival of the oscillation is a
purely quantum phenomenon and has no simple quasiclassical explanation. For
more details, see \cite{RSS2009}.

\subsection{\it Observation of bound quantum states of neutrons in the gravitational field of the earth\label{sec:boundqs}}

\begin{figure}[tb]
\subfigure{\hspace{2.5cm}\epsfig{file=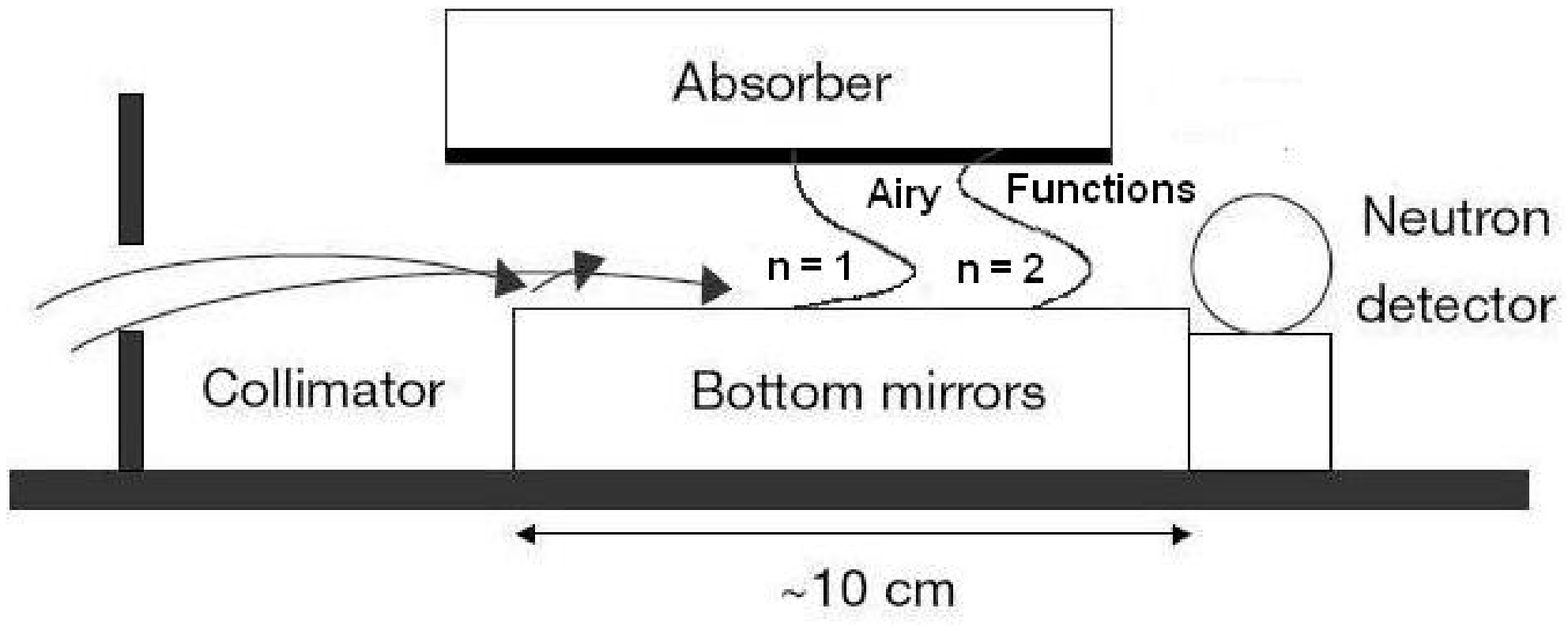,scale=.55}}
\caption{Sketch of the setup of the first observation of bound quantum states
in the gravitational field of the earth. Mirror and absorber system with
adjustable slit height are shown.
\label{gravitysetup}}
\end{figure}

The first observation of quantum states in the gravitational potential of the
earth with ultracold
neutrons~\cite{Nesvizhevsky02,Westphal06,Nesvizhevsky03,Nesvizhevsky05} has
been performed at the Institut Laue-Langevin and started as a
collaboration between ILL (Grenoble), PNPI (Gatchina), JINR (Dubna) and the
Physics Institute at Heidelberg University. In this experiment neutrons were
allowed to populate the lower states of Fig.~\ref{fig:states}. Higher,
unwanted states were removed. The neutrons pass through the
mirror-absorber/scatterer-system shown in Fig.~\ref{gravitysetup} and are
eventually detected by a $^3$He-counter. The height of the absorber above the
mirror is varied and the transmission is measured as a function of height.

\begin{figure}[tb]
\subfigure{\hspace{2.0cm}\epsfig{file=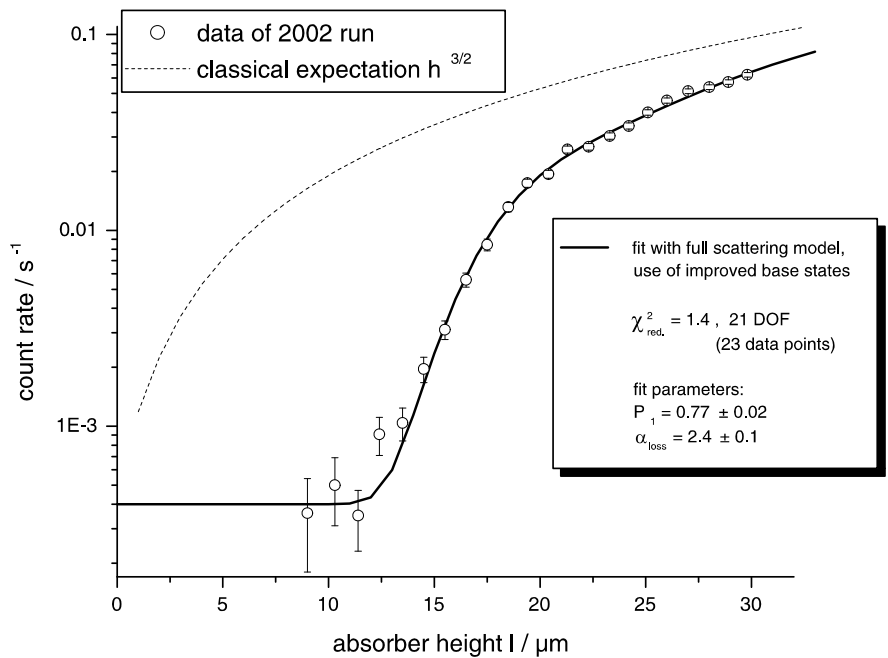,scale=.45}}
\caption{Data 2002~\cite{Nesvizhevsky02} in logarithmic scale: A fit to the transmission as a function of the
absorber height $h$ = $l$~\cite{Westphal06} The difference between the quantum observation and the classical expectation is clearly visible.
\label{gravity2002}}
\end{figure}

\begin{figure}[tb]
\begin{center}
\begin{minipage}[t]{8 cm}
\hspace{-2cm}\epsfig{file=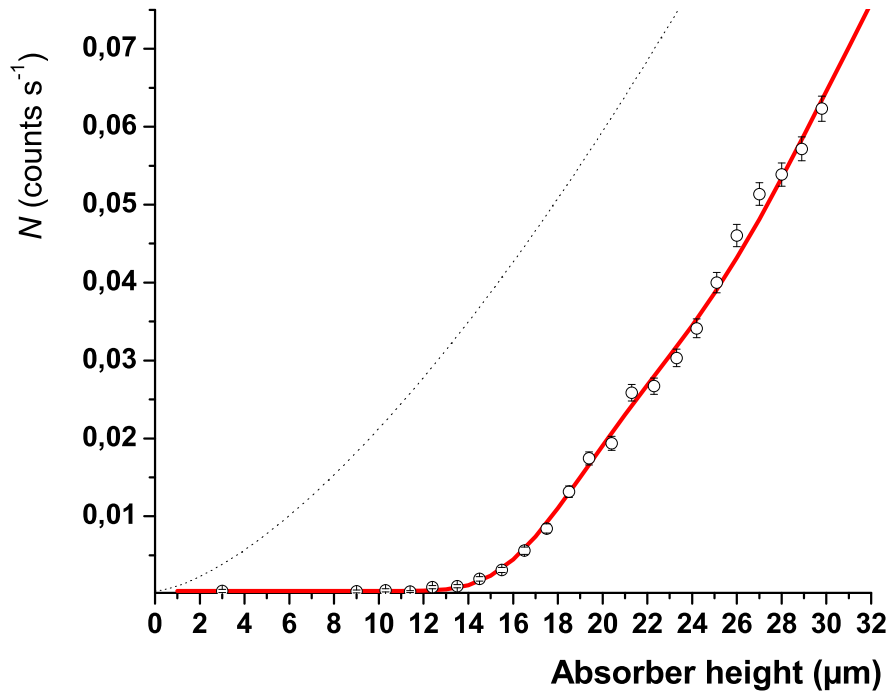,scale=1.25}
\end{minipage}
\begin{minipage}[t]{15. cm}
\caption{Data 2005~\cite{Nesvizhevsky05} in linear scale: A fit to the transmission as a function of the
absorber height $l$~\cite{Westphal06}. The dotted
line shows the classical expectation.\label{online}}
\end{minipage}
\end{center}
\end{figure}

The data points, plotted in Fig. \ref{gravity2002} and Fig.~\ref{online} show
the measured number of transmitted neutrons for an absorber height $l$ and
follow the expected behavior: No neutrons reach the detector below an absorber
height of $15\mu$m. Then, above an absorber height of 15 $\mu$m, one sees the
transmission of ground-state neutrons resulting in an increased count
rate. From the classical point of view, the transmission $F$ of neutrons is
proportional to the phase space volume allowed by the absorber. It is governed
by a power law $F$ $\sim$ $l^n$ and $n = 3/2$. The dotted line shows this
classical expectation. The difference between the quantum observation and the
classical expectation is clearly visible.

A description of the measured transmission $F$ versus the absorber height $l$
can be obtained via the solution of the decoupled one-dimensional stationary
Schr\"odinger equation in Eq.~(\ref{eq:sgl}). The general solution of
Eq.~(\ref{eq:sgl}) for $\psi_n$ is a superposition of Airy functions $Ai(z)$ and $Bi(z)$,
\begin{equation}
\psi_{n}(z)=A_n\cdot Ai\left(\zeta-\zeta_n\right)+B_n\cdot
Bi\left(\zeta-\zeta_n\right) \label{eq:aibi}
\end{equation}
with the scaling $\zeta$ and a displacement $\zeta_E$ of Eq.~(\ref{eq:zeta}). There is an additional boundary condition at the absorber and
we need the term with Airy's $Bi$-function going to infinity at
large $z$. Due to the absorber, the bound neutron
states $\psi_{n}$ in the linear gravitational potential are confined by two
very high potential steps above (absorber) and below (mirror). In the case of
$l\leq z_n$, the absorber/scatterer will begin to change the bound
states. This implies an energy shift leading to $E_{n}(l)\gg mg\cdot
l$~\cite{Westphal02}. Therefore, the calculation of the loss rate is performed
using the full realistic bound states of Eq.~(\ref{eq:aibi}). Only for large $l$, the n-th eigenfunction of the neutron mirror system reduces to
the Airy function $Ai(\zeta -\zeta_n)$ for $n>0$. The corresponding
eigenenergies are $E_n=m_ngz_n$ with $z_n = z_o\zeta_n$. In the
WKB-approximation one obtains in leading order
\be
\zeta_n=\left(\frac{3\pi}{2}(n-\frac{1}{4})\right)^{2/3}\, ,
\end{equation}
where $z_n$ corresponds to the turning point of a classical neutron trajectory
with energy $E_n$.

In a different setup, the absorber was placed at the bottom and the
mirror on top. Using this reversed geometry, the scattering-inducing
roughness is found at $z=0$. The bound states are now, again, given
by the Airy functions, but the gravitationally bound states will be
strongly absorbed at arbitrary heights $l$ of the mirror at top --
in contrast to the normal setup. Thus, the dominating effect of
gravity on the formation of the bound states has been demonstrated,
since the measurement did not just show a simple confinement
effect~\cite{Westphal06}.

\section{The gravity resonance spectroscopy method}
A previously developed resonance spectroscopy technique~\cite{Jenke2011} allows a precise measurement~\cite{Abele2010} of the energy levels of quantum states in the gravity potential of the earth. Quantum mechanical transitions with a characteristic energy exchange between an
externally driven modulator and the energy levels are observed on resonance.
An essential novelty of this kind of spectroscopy is the fact that the quantum
mechanical transition is mechanically driven by an oscillating mirror and is not
a consequence of a direct coupling of an electromagnetic charge or moment to an
electromagnetic field. On resonance energy transfer is observed according to the
energy difference between gravity quantum states coupled to the modulator. The physics behind these transitions is related to earlier studies of energy transfer when matter waves bounce off a vibrating mirror~\cite{Hamilton1987,Felber1996} or a time-dependent crystal~\cite{Steane1995,Bernet1996}.

The concept is related to Rabi's magnetic resonance technique for measurements of nuclear magnetic moments~\cite{1}. The sensitivity is extremely high, because a quantum mechanical phase shift is converted into a frequency measurement. The sensitivity of resonance methods reached so far~\cite{17} is $6.8 \times 10^{-22}$ eV and has been achieved in a search for a non-vanishing electric dipole moment of the neutrons. Such an uncertainly in energy corresponds to a time uncertainty of six days according to Heisenberg's principle. \\

In a two-level spin-$1/2$-system coupled to a resonator, magnetic transitions occur, when the oscillator frequency $\omega$ equals the Bohr frequency of the system. Rabi resonance spectroscopy measures the energy difference between these levels $\left|p\right\rangle$ and $\left|q\right\rangle$ and damping $\gamma$. The wave function of the two level system is
\be
\Psi \left(\bar r ,t \right) =  \left\langle\bar{r}|\Psi\left(t\right)\right\rangle = C_p\left(t\right)e^{-i\omega_p t}u_p \left(\bar r\right) + C_q \left(t\right) e^{-i\omega_q t}u_q\left(\bar r\right)
\ee
with the time varying coefficients $C_p\left(t\right)$ and $C_q\left(t\right)$.\\
With the frequency difference $\omega_{pq}$ between the two states, the frequency $\omega$ of the driving field, the detuning $\delta\omega = \omega_{pq} - \omega$, the Rabi frequency $\Omega_R$ and the time $t$, the coupling between the time varying coefficients is given by
\be
\frac{d}{dt}\left( \begin{array}{c} \tilde{C_p}\left(t\right) \\ \tilde{C_q}\left(t\right) \end{array} \right) = \frac{i}{2}\left(\begin{array}{cc} \delta\omega & \Omega_R \\ \Omega_R^* & -\delta\omega  \end{array} \right) \left( \begin{array}{c} \tilde{C_p} \\ \tilde{C_q} \end{array} \right)
\ee

with a transformation into the rotating frame of reference:
 \be
 \begin{array}{cc}
  C_p\left(t\right) &= \tilde C_p\left(t\right)\cdot e^{-\frac{i}{2}\delta \omega t}\\
  C_q\left(t\right) &= \tilde C_q\left(t\right)\cdot e^{-\frac{i}{2}\delta \omega t}.
  \end{array}
 \ee
Here, $\Omega_R$  is a measure of the strength of the coupling between the two levels and is related to the vibration strength.
Such oscillations are damped out and the damping rate depends on how strongly the system is coupled to the environment.

In generalizing this system, it is possible to describe quantum states in the gravity field of the Earth in analogy to a spin-$1/2$-system, where the time development is described by the Bloch equations. Neutron matter waves are excited by an oscillator coupled to quantum states with transitions between state $\left|p\right\rangle$ and state $\left|q\right\rangle$. The energy scale is the pico-eV-scale.

In the experiment, the damping is caused by the scatterer \cite{Westphal06} at height $l$ above a mirror. A key point for the demonstration of this method is that it allows for the detection of resonant transitions $|p\rangle \rightarrow |q\rangle$ at a frequency, which is tuned by this scatterer height. As  explained, the additional mirror potential shifts the energy of state $|3\rangle$ as a function of height, see Fig. \ref{fig:2}. The absorption is described phenomenologically by adding decay terms $\gamma_p$  and $\gamma_q$ to the equations of motion:

\be
\frac{d}{dt}\left( \begin{array}{c} \tilde{C_p}\left(t\right) \\ \tilde{C_q}\left(t\right) \end{array} \right) = \frac{i}{2}\left(\begin{array}{cc} \delta\omega + i\gamma_p & \Omega_R \\ \Omega_R^* & -\delta\omega+i\gamma_q  \end{array} \right) \left( \begin{array}{c} \tilde{C_p} \\ \tilde{C_q} \end{array} \right)
\ee

The oscillator is realized by a vibrating mirror i.e. a modulation of the hard surface potential in vertical position. Neutron mirrors are made of polished optical glass. Interactions limit the lifetime of a state of a two-level system. Lifetime limiting interactions are described phenomenologically by adding decay terms to the equations of motion leading to damped oscillation. Another concept has been proposed by~\cite{21}.

\begin{figure}[tb]
\subfigure{\hspace{.5cm}\epsfig{file=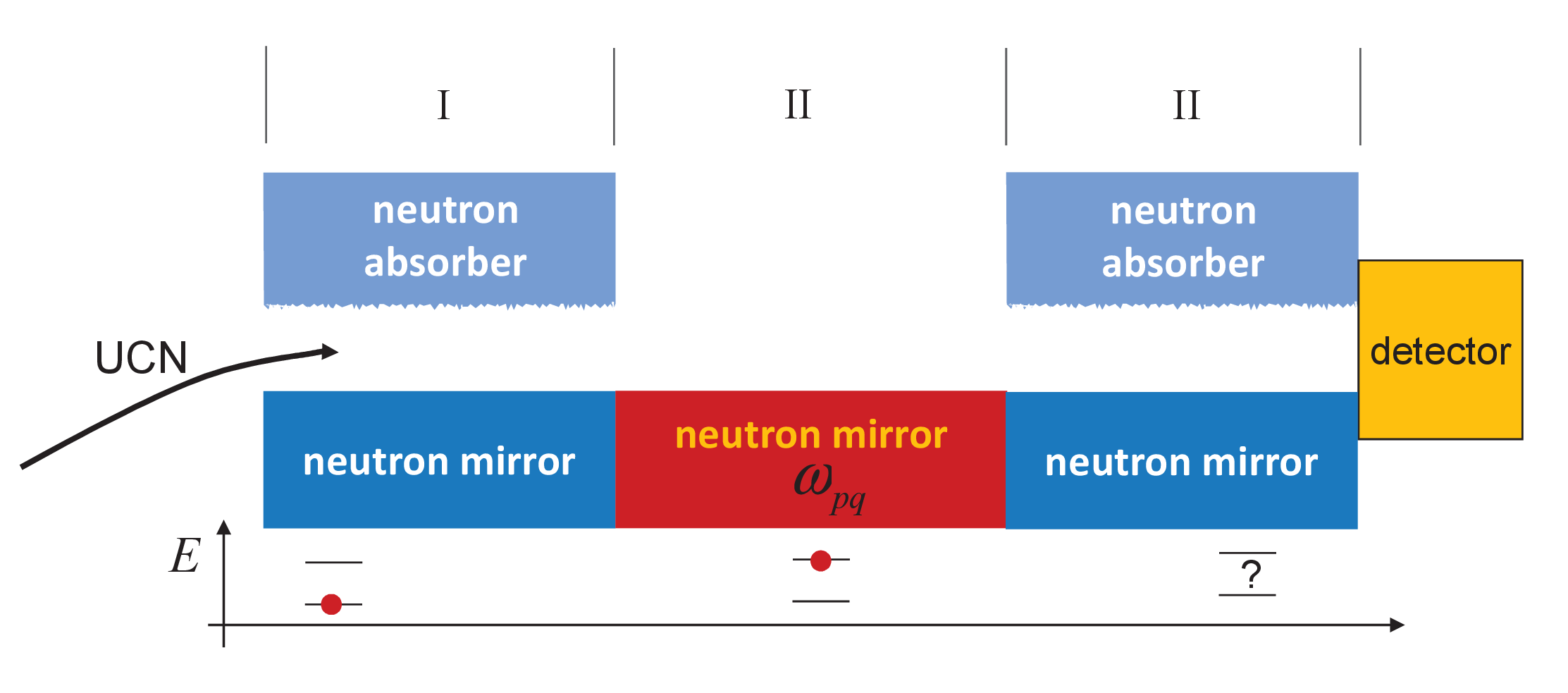,scale=.44}}\hspace{.5cm}\epsfig{file=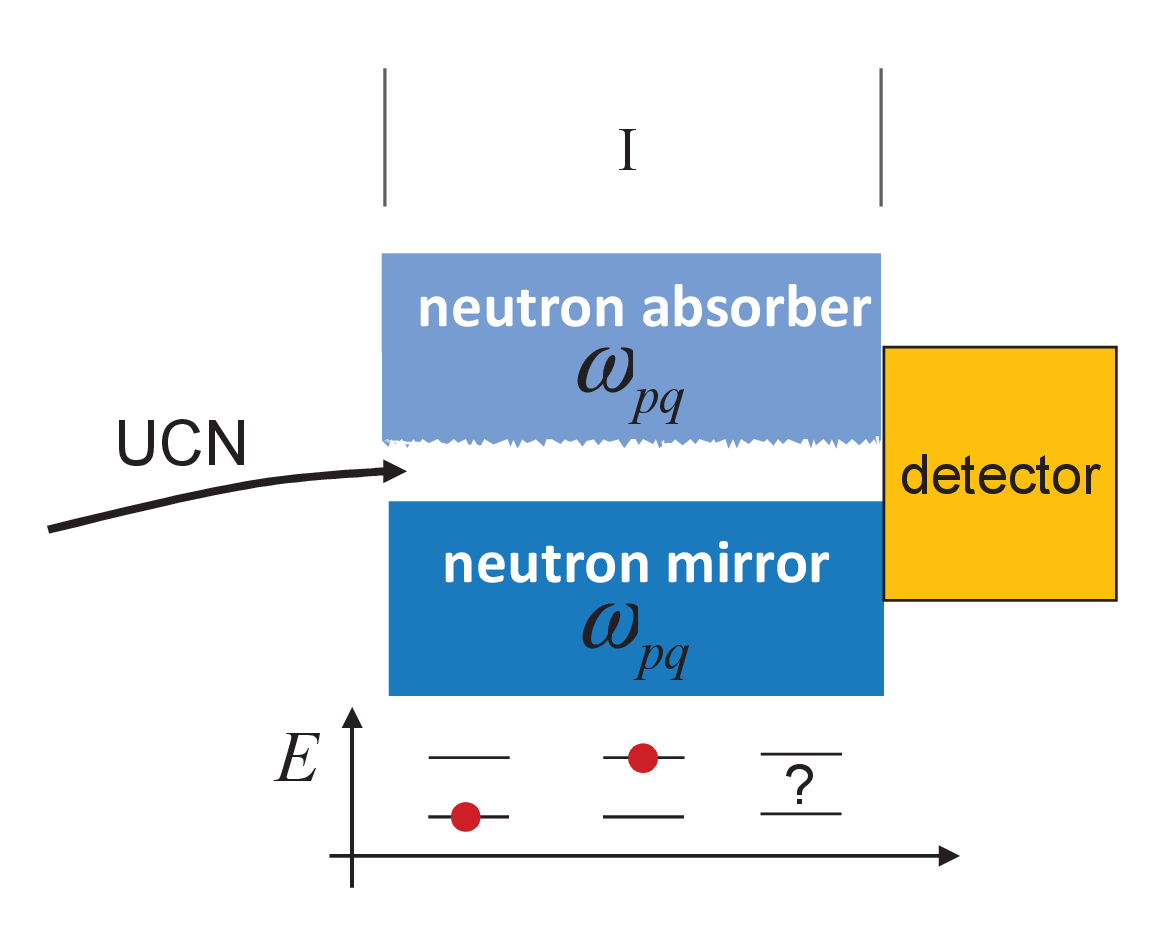,scale=.46}
\caption{Comparison of two different Resonance Spectroscopy Techniques~\cite{Jenke2011}}
\label{fig:rabi}
\end{figure}

 A typical Rabi resonance spectroscopy experiment consists of three regions, where particles pass through. One firstly has to install a state selector in region~I, secondly a coupling region inducing specific transitions in the two state systems, whose energy difference is measured in region~II, and a state detector in region~III, see Fig. \ref{fig:rabi}, left. A so called $\pi$-pulse in region~II creates an inversion of the superposition of the two states, whose energy difference is to be measured. Technical details are described in~\cite{Abele2010}.

 In this experiments with neutrons, regions~I to III are achieved with only one bottom mirror coupled to a mechanical oscillator, a scatterer on top and a neutron detector behind, see Fig. \ref{fig:rabi}, right. The scatterer only allows the ground state to pass and prepares the state $\left|p\right\rangle$. It removes and absorbs higher, unwanted states~\cite{Westphal06} as explained. The vibration, i.e. a modulation of the mirror's vertical position, induces transitions to $\left|q\right\rangle$, which are again filtered out by the scatterer. The neutrons are taken from the ultra-cold neutron installation PF2 at Institute Laue-Langevin. The horizontal velocity is restricted to $5.7 \mbox{m/s} < v < 7 \mbox{m/s}$. The experiment itself is mounted on a polished plane granite stone with an active and a passive anti-vibration table underneath. This stone is leveled with a precision better than 1 $\mu$rad. A mu-metal shield suppresses the coupling of residual fluctuations of the magnetic field to the magnetic moment of the neutron is sufficiently.\\

\subsection{\it Experimental results}
\begin{figure}[!htb]
\subfigure{\hspace{1.cm}\epsfig{file=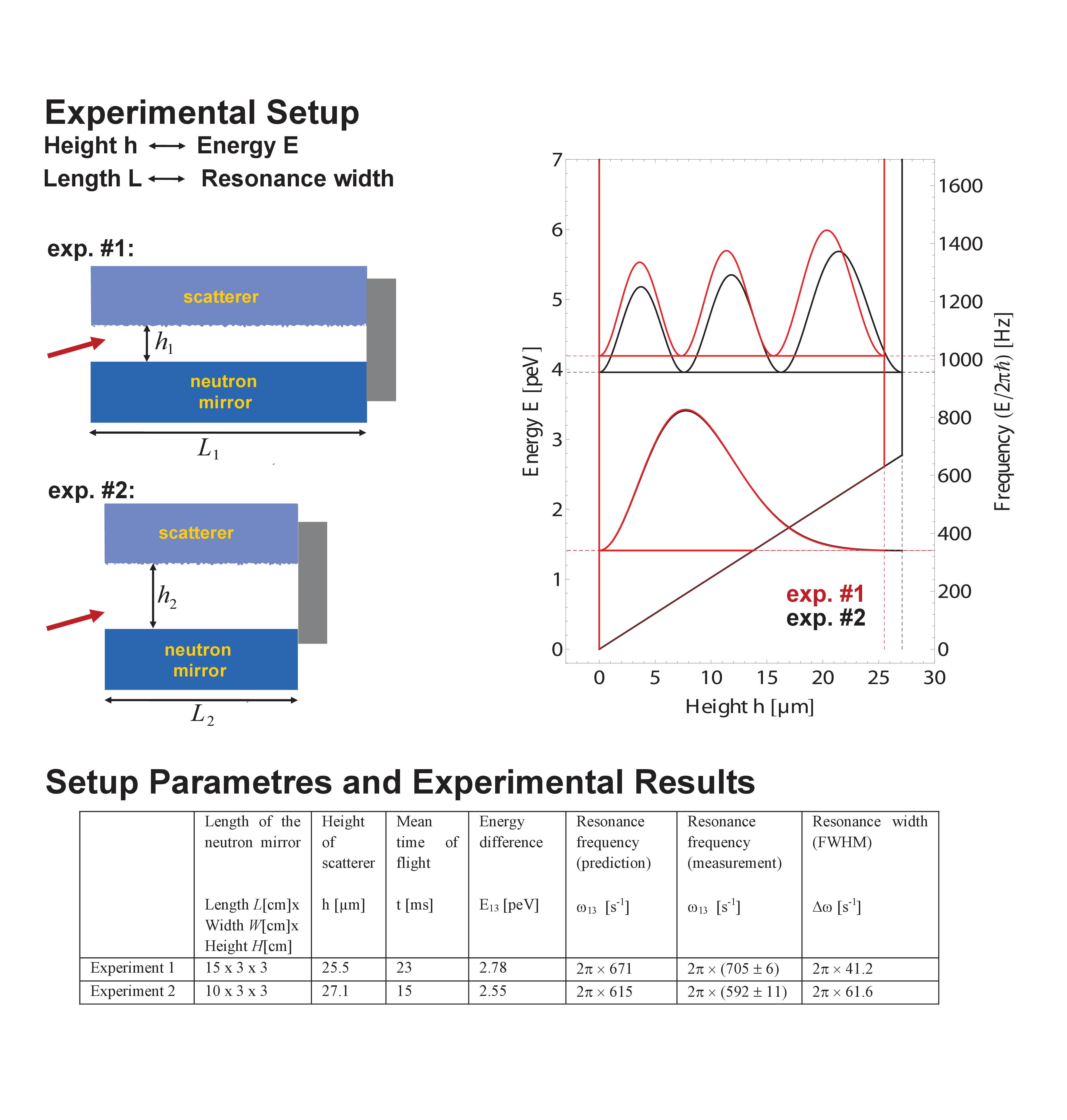,scale=.35}}
\caption{Experimental Parameters (Figure taken from~\cite{Jenke2011}).}
\label{fig:2}
\end{figure}

\begin{figure}[!htb]
		 \includegraphics[width=0.75\textwidth\hspace{2.cm}]{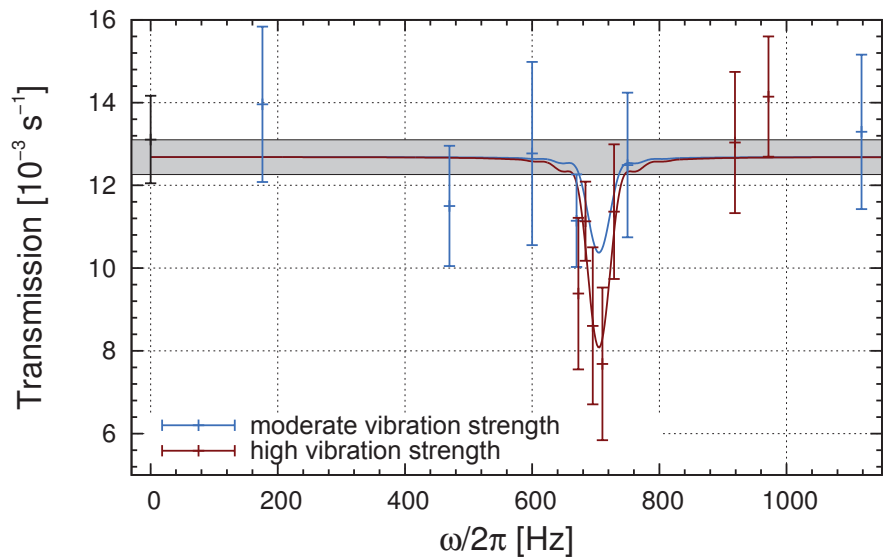}
			
\caption{Gravity resonance spectroscopy and excitation~\cite{Jenke2011}: The transmission as a function of modulation frequency shows a sharp resonance at $\omega_{13} = 2\pi \times (705 \pm 6) \mbox{s}^{-1}$. The grey band represents the statistical $1\sigma$ uncertainty of all off-resonant data points. Blue (brown) data points correspond to measurements with moderate (high) vibration strength $1.5 \leq a \leq 4.0 \mbox{m/s}^2 (4.9 \leq a \leq 7.7 \mbox{m/s}^2)$. The corresponding Rabi resonance curve is calculated using their mean vibration strength of $2.95 \mbox{m/s}^2 (5.87 \mbox{m/s}^2)$. The black data point sums up all measurements at zero vibration.}\label{fig:np1}
\end{figure}

\begin{figure}
\includegraphics[width=0.75\textwidth\hspace{2.cm}]{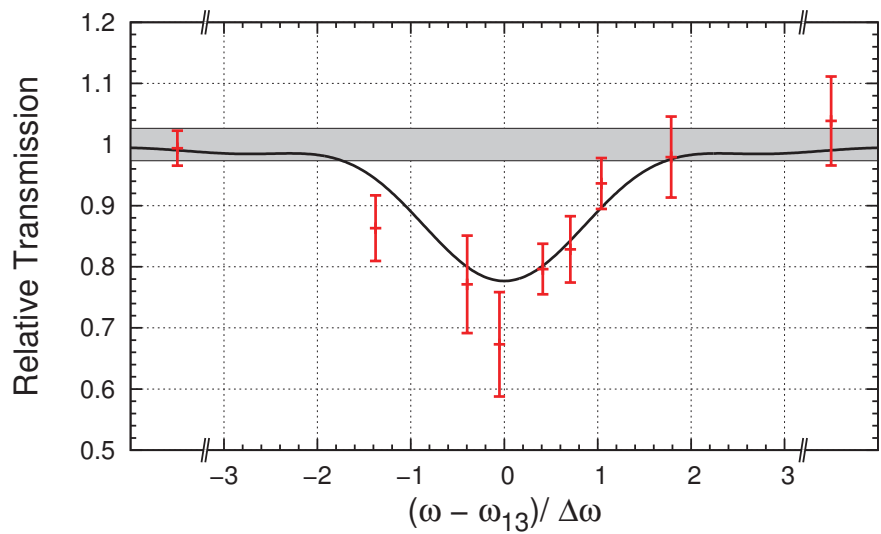}
			
		\caption{Combined result for both measurements with mirror length $L = 10$ cm and $L = 15$ cm. The transmission in units of the unperturbed system is displayed as a function of detuning. The significance for gravity spectroscopy between state $\left|1\right\rangle$ and $\left|3\right\rangle$ at $\omega_{13}$ is 4.9 standard deviations. The left and right data points combine all off-resonant measurements with $\left|\left(\omega - \omega_{13}\right)/\Delta\omega\right|\geq3$.}\label{fig:np2}
\end{figure}

Within the qBounce experiment~\cite{9}, several resonance spectroscopy measurements with different geometric parameters have been performed, resulting in different resonance frequencies and widths. In general, the oscillator frequency at resonance for a transition between states with energies $E_p$ and $E_q$ is
\begin{equation}
\omega_{pq}=\frac{E_q-E_p}{\hbar}=\omega_q-\omega_p.
\end{equation}
The transfer is referred to as Rabi transition. The transitions $|1\rangle \leftrightarrow |2\rangle$, $|1\rangle \leftrightarrow |3\rangle$, $|2\rangle \leftrightarrow |3\rangle$, and $|2\rangle \leftrightarrow |4\rangle$ have been measured.
In detail, the $|1\rangle \leftrightarrow |3\rangle$ transition with $\omega_{13} = \omega_3 - \omega_1$ is described (see table in Fig.~\ref{fig:2}).
On resonance ($\omega = \omega_{13}$), this oscillator drives the system into a coherent superposition of state $\left|1\right\rangle$ and $\left|3\right\rangle$ and one can choose the amplitude $a$ in such a way that we have complete reversal of the state occupation between $\left|1\right\rangle$ and $\left|3\right\rangle$. It is - as it has been done - convenient to place the scatterer at a certain height $h$ on top of the bottom mirror. This allows to tune the resonance frequency between $\left|1\right\rangle$ and $\left|3\right\rangle$ due to the additional potential of the scatterer, which shifts the energy of state $\left|2\right\rangle$ and $\left|3\right\rangle$, but leaves state $\left|1\right\rangle$ unchanged, see Fig.~\ref{fig:2}. The energy levels and the probability density distributions for these states are also given in Fig. \ref{fig:2}. The scatterer removes neutrons from the system and the Rabi spectroscopy contains a well defined damping.\\
The observable is the measured transmission $\left|1\right\rangle$ to $\left|3\right\rangle$ as a function of the modulation frequency and amplitude, see Fig. \ref{fig:np1} and \ref{fig:np2}. For this purpose, piezo-elements have been mounted underneath. They induce a rapid modulation of the surface height with acceleration $a$, which was measured with a noise and vibration analyzer attached to the neutron mirror system. In addition to this, the position-dependent mirror vibrations were measured using a laser-based vibration analyzer attached to the neutron mirror systems.\\

For the first experiment, Fig. \ref{fig:np1} shows the measured count rate as a function of $\omega$. Blue (brown) data points correspond to measurements with moderate (high) vibration strength \mbox{$1.5 \leq a \leq 4.0\ \mbox{m/s}^2\ (4.9 \leq a \leq 7.7\ \mbox{m/s}^2)$}. The corresponding Rabi resonance curve was calculated using their mean vibration strength of $2.95 \mbox{m/s}^2 (5.87 \mbox{m/s}^2)$. The black data point sums up all of the measurements at zero vibration. The gray band represents the one sigma uncertainty of all off-resonant data points. The brown line is the quantum expectation as a function of oscillator frequency $\omega$ for Rabi transitions between state $\left|1\right\rangle$ and state $\left|3\right\rangle$ within an average time of flight $\tau=L/v=23$ ms. The normalization for transmission $T$, frequency at resonance $\omega_{13}$ and global parameter $f$ are the only fit parameters. It was found that the vibration amplitude does not change on the flight path of the neutrons but depends in a linear way on their transversal direction. $f$ is a weighting parameter to be multiplied with the measured vibration strength to correct for these linear effects. A sharp resonance was found at frequency $\omega_{13}=2\pi \times (705 \pm 6)$~Hz, which is close to the frequency prediction of $\omega_{13}=2\pi \times 671$ Hz, if we remember that the height measurement has an uncertainty due to the roughness of the scatterer. For the weighting factor we found $f = 0.56 \pm 0.16$. The full width at half maximum is the prediction made from the time the neutrons spend in the modulator. The significance for $\left|1\right\rangle \rightarrow \left|3\right\rangle$ excitations is 3.5 standard deviations.\\


The fit used in Fig.~\ref{fig:np1} contains three parameters, the resonant frequency $\omega_{pq}$, the transmission normalization $N$, and vibration strength parameter $f$ to be multiplied with the measured acceleration. The damping as a function of $\Omega_R$ was measured separately, see Fig.~\ref{fig:np2}, as well as the width of the Rabi oscillation and the background ($0.005 \pm 0.0002\ \mbox{s}^{-1}$).\\

In a second measurement, the length $L = 10$ cm reduces the average flight time to $\tau = 15$ ms. Furthermore, the scatterer height differs by 1.6 $\mu$m from the first measurement, thus changing the resonant frequency prediction to $\omega_{13} = 2\pi \times  615$ Hz. The resonance frequency $\omega_{13} = 2\pi \times (592 \pm 11)$ Hz is close to the prediction and $f = 0.99 \pm 0.29$ are observed. Fig.~\ref{fig:np2} shows the combined result for both measurements with mirror length $L$ = 10 cm and $L$ = 15 cm. The transmission in units of the unperturbed system is displayed as a function of detuning. In total, the significance for gravity resonance spectroscopy between state $\left|1\right\rangle$ and $\left|3\right\rangle$ at $\omega_{13}$ is 4.9 standard deviations. The left and right data points combine all off-resonant measurements with $\left|\left(\omega - \omega_{13}\right)/\Delta\omega\right|\geq3$, where $\Delta \nu$ is the half width at half maximum.

\section{Summary and outlook}
Gravity experiments with neutrons are motivated mainly due to the fact that in contrast to atoms the electrical
polarizability of neutrons is extremely small. Neutrons are not
disturbed by short range electric forces such as van der Waals or
Casimir forces. Together with its neutrality, a neutron provides a sensitivity many orders of magnitude below the strength of
electromagnetism.

"For the experiments
discussed, the preliminary estimated sensitivity of the measured energy difference between the gravity levels is $7.6 \times 10^{-3}$. This corresponds to $\Delta E = 2 \times 10^{-14}$ eV, which tests Newton's law on the micrometre distance at this level of precision.\\
This accuracy range is of
potential interest, because it addresses some of the unresolved questions
of science: the nature of the fundamental forces and under­lying symmetries and the nature of gravitation at small distances~\cite{10}. Hypothetical extra-dimensions, curled up to cylinders or tori with a small compactification radius would lead to deviations from Newton's gravitational law at very small distances~\cite{9}. Another example is suggested by the magnitude of the vacuum energy in the universe~\cite{24,25}, which again is linked to the modification of gravity at small distances. Furthermore, the experiments have the potential to test the equivalence principle~\cite{Kajari2010}.\\

The long term plan is to apply Ramsey's method of separated oscillating fields to the spectroscopy of the quantum states in the gravity potential above a horizontal mirror~\cite{Abele2010}. Such measurements with ultra-cold neutrons will offer a sensitivity to Newton's law or hypothetical short-ranged interactions that is about 21 orders of magnitude below the energy scale of electromagnetism.

\section*{Acknowledgments}
{\footnotesize
We gratefully acknowledge support from the Austrian Science Fund (FWF) under Contract No. I529-N20 and  the German Research Foundation (DFG) within the Priority Programme (SPP) 1491 "Precision experiments in particle and astrophysics with cold and ultracold neutrons", the DFG Excellence Initiative "Origin of the Universe", and DFG support under Contract No. Ab128/2-1.
}


\section*{References}
\begin{thebibliography}{10}
\def\ARNP{{\em  Ann. Rev. Nucl. Part. Sci}}
\def\JETP{{\em JETP}}
\def\SPJETP{{\em JETP}}
\def\SJNP{{\em Sov. J. Nucl. Phys.}}
\def\IJMPA{{\em Int. J. Mod. Phys.} A}
\def\JETPL{{\em JETPL}}
\def\NIMA{{\em Nucl. Instrum. Meth. Phys. Res.,} Sec. A}
\def\NIMB{{\em Nucl. Instrum. Meth. Phys. Res.,} Sec. B}
\def\PR{{\em JETPL}}
\def\EPL{{\em Europhys. Lett.}}
\def\EPJA{{\em Euro. Phys. J} A}
\def\EPJC{{\em Euro. Phys. J} C}
\def\JPG{{\em J. Phys.} G}
\def\Can. J. Phys.{{\em Can. J. Phys.}}
\def\NC{{\em Nuovo Cimento}}
\def\PPNP{{\em Progress in Particle and Nuclear Physics}}
\def\Contemp. Phys.{{\em Contemp. Phy.}}
\def\PREP{{\em Phy. Rep.}}
\def\PHYSB{{\em Physica} B}

\itemsep -2pt
{\footnotesize
\bibitem{Colella75}R. Collela, A.W. Overhauser, and S.A. Werner, \J{\PRL}{34}{1472}{1975}.

\bibitem{Jenke09}T. Jenke et al., Nucl. Instrum. Meth. A {\bf 611} 318 2009.

\bibitem{Abele09}H. Abele et al., Nuclear Physics {\bf A827} 593c (2009).

\bibitem{RSS2009}R. Reiter, B. Schlederer, D. Seppi, {\em Quantenmechanisches
Verhalten eines ultrakalten Neutrons im Gravitationsfeld}, Projektarbeit,
Atominstitut, Vienna University of Technology, Vienna, 2009, unpublished.
\bibitem{Nesvizhevsky02}V. Nesvizhevsky et al., \J{\em Nature}{415}{297}{2002}.
\bibitem{Jenke2011}T. Jenke et al., \textit{Nature Physics} \textbf{7}, 468 (2011).
\bibitem{Durstberger2011}K. Durstberger et al., \textit{Phys. Rev.} \textbf{D}, 84 (2011) 036004.

\bibitem{Abele2010}Abele, H. et al., \textit{Phys. Rev.} \textbf{D81}, 065019 (2010).
\bibitem{AbelePPNP}H. Abele, Progress in Particle and Nuclear Physics {\bf 60} 1 (2008).
\bibitem{Rauch2011}Hasegawa Y and Rauch H, 2011 New Journal of Physics 13 115010
\bibitem{Rauch00} H. Rauch and S. A. Werner, {\it Neutron
Interferometry, Lessons in Experimental Quantum Mechanics} (Oxford
Science Publications, Clarendon Press, Oxford 2000).
\bibitem{Byrne93}J. Byrne, \emph{Neutrons, Nuclei and Matter, An
Exploration of the Physics of Slow Neutrons}, (Institute of Physics
Publishing Bristol and Philadelphia, 1993).
\bibitem{Aminoff93}C.G.Aminoff et al., \J{\PRL}{71}
{3083}{1993}.
\bibitem{Kasevich90} M.A. Kasevich, D.S. Weiss, and S. Chu, \J{\em Opt.
Commun.}{15}{607}{1990}.
\bibitem{Roach95}T.M. Roach et al., \J{\PRL}{75}{629}{1995}.
\bibitem{McReynolds51}A.W. McReynolds, \J{\PR}{83}{233}{1951}.
\bibitem{Dabbs65}J.W.T. Dabbs et al., \J{\PR}{139}{765}{1965}.
\bibitem{Koester76}L. Koester \J{\PRD}{14}{907}{1976}.
\bibitem{Sears82}V.F. Sears, {\em Phys. Rep.} 82 (1982) 1.

\bibitem{Rauch74}H. Rauch, W. Treimer, H. Bonse, \J{\PLA}{47}{369}{1974}.
\bibitem{Werner88}S. A. Werner et al., \J{\em Physica B}{151}{22}{1988}.
\bibitem{Littrell97} K.C. Littrell et al., \J{\PRA}{56}{1767}{1997}.
\bibitem{Werner86}S.A. Werner, A.G. Klein, \J{\em Meth. Exp. Phys.}{23A}{259}{1986}.
\bibitem{Atwood84}D.K. Atwood et al., \J{\PL}{52}{1673}{1984}.
\bibitem{Staudenmann80}J.L. Staudenmann et al., \J{\PRA}{21}{1419}{1980}.
\bibitem{Zouw00}G. van der Zouw et al., \J{\NIM}{A440}{568}{2000}.
\bibitem{Mannheim98}P.D. Mannheim, \J{\PRA}{57}{1260}{1998}.
\bibitem{Cohen93}J.M. Cohen and B. Mashhoon, \J{\PLA}{181}{353}{1993}.
\bibitem{Pritchard2009}A. D. Cronin, J. Schmiedmayer, D. E. Pritchard, {\em Rev. Mod. Phys.} 81 (2009] 1051.
\bibitem{Peters99}A. Peters, K.Y. Chung, and S. Chu, \J{\em Nature (London)}{400}{849}{1999}.
\bibitem{Mueller2010}H. M\"uller, A. Peters, S. Chu, \J{\em Nature}{463}{927}{2010}.
\bibitem{Mueller2010a}H. M\"uller, A. Peters, S. Chu, \J{\em Nature}{467}{E2}{2010}.
\bibitem{Wolf2010}P. Wolf et al., \J{\em Nature}{467}{E1}{2010}.
\bibitem{Wolf2011}P. Wolf et al., \{\em Class. Quantum Grav.}{28}{145017}{2011}.



\bibitem{Gibbs}R.~L. Gibbs, Am. J. Phys. {\bf 43} 25 (1975).
\bibitem{Rosu}H.~C. Rosu, [arXiv:quant-ph/0104003].
\bibitem{qqb}Julio Gea-Banacloche, Am. J. Phys. {\bf 67} 776 (1999).

\bibitem{Wallis}H. Wallis et. al., Appl. Phys. B 54 407 (1992).


\bibitem{Lushikov}V.I. Luschikov and A.I. Frank, JETP Lett. 28 559
(1978).
\bibitem{Westphal02}A. Westphal, Diploma thesis, University of Heidelberg (2001), arXiv:gr-qc/0208062.
\bibitem{Voronin06}A.Yu. Voronin et al., \J{\PRD}{73}{044029}{2006}.
\bibitem{Westphal06}A. Westphal et al., Eur. Phys. J \textbf{C51}, 367 (2007).arXiv:hep-ph/0602093.
\bibitem{Ruess00}T. Ruess, Diploma thesis, University of Heidelberg, unpublished (2000).

\bibitem{Nesvizhevsky03}V. Nesvizhevsky et al., \J{\PRD}{67}{102002}{2003}.
\bibitem{Nesvizhevsky05}V.V. Nesvizhevsky et al., \J{\EPJC}{40}{479}{2005}.
\bibitem{1}Rabi I. et al., \textit{Phys. Rev}. \textbf{55}, 526 (1939).
\bibitem{17}Baker, C. A. et al., \textit{Phys. Rev. Lett.} \textbf{97}, 131801 (2006).
\bibitem{21}Kreuz, M. et al., arXiv:physics.ins-det 0902.0156.
\bibitem{Hamilton1987}W. A. Hamilton et al., \textit{Phys. Rev. Lett.} \textit{58}, 2770 (1987).
\bibitem{Felber1996}J. Felber et al,  \textit{Phys. Rev.} \textit{A53}, 319 (1996).
\bibitem{Steane1995}A. Steane et al., \textit{Phys. Rev. Lett.} \textit{74}, 4972 (1995).
\bibitem{Bernet1996}S. Bernet et al., \textit{Phys. Rev. Lett.} \textit{77}, 5160 (1996).

\bibitem{9}Arkani-Hamed, N., Dimopolos, S. \& Dvali, G., \textit{Phys. Rev.} \textbf{D59}, 086004 (1999).
\bibitem{10}Fischbach E. \& Talmadge C. L. The search for non-Newtonian gravity. Springer-Verlag New York (1999).
\bibitem{24}Callin, P. \& Burgess, C. P., \textit{Nucl.Phys.} \textbf{B752} 60-79 (2006).
\bibitem{25}Sundrum, R., \textit{J. High Energy Phys.} \textbf{07}, 001 (1999).
\bibitem{Kajari2010}E. Kajari et al., \textit{Appl. Phys.} \textit{B 100}, 43-60 (2010).
\bibitem{Jenke2011}T. Jenke, Dissertation Thesis 2011, Vienna University of Technology unpublished.







\end{thebibliography}
\end{document}